\documentclass[a4paper,onecolumn,11pt,unpublished]{quantumarticle}
\pdfoutput=1
\usepackage[utf8]{inputenc}
\usepackage[english]{babel}
\usepackage[T1]{fontenc}
\usepackage{amsmath,amssymb,amsfonts}
\usepackage[numbers,sort&compress]{natbib}
\usepackage{hyperref}
\usepackage{mathtools}
\usepackage{dsfont}
\usepackage[shortlabels]{enumitem}
\usepackage{bm}
\usepackage[mathscr]{eucal}

\newtheorem{theorem}{Criterion}
\newtheorem{definition}{Definition}

\newcommand{\objQ}[1]
		{ q_E^{(#1)} }					
\newcommand{\objP}[1]
		{ p_E^{(#1)} }					
		
\newcommand{\boldPrm}
		{ \bm{\xi} }
\newcommand{\boldBs}
		{ \bm{\zeta} }
\newcommand{\prm}
		{ \xi }
\newcommand{\bs}
		{ \zeta }

\newcommand{\boldT}
		{ \mathbf{t} }
		
\newcommand{\dMap}[1]
		{ \Lambda(#1)}
		
\usepackage{ulem}

\begin{document}

\title{Noise representations of open system dynamics}

\author{Piotr Sza\'{n}kowski}\email{piotr.szankowski@ifpan.edu.pl}
\affiliation{Institute of Physics, Polish Academy of Sciences, al.~Lotnik{\'o}w 32/46, PL 02-668 Warsaw, Poland}
\orcid{0000-0003-4306-8702}
\author{{\L}ukasz Cywi\'{n}ski}\email{lcyw@ifpan.edu.pl}
\affiliation{Institute of Physics, Polish Academy of Sciences, al.~Lotnik{\'o}w 32/46, PL 02-668 Warsaw, Poland}
\orcid{0000-0002-0162-7943}

\begin{abstract}
We analyze the conditions under which the dynamics of a quantum system open to a given environment can be simulated with an external noisy field that is a surrogate for the environmental degrees of freedom. We show that such a field is either a subjective or an objective surrogate; the former is capable of simulating the dynamics only for the specific system--environment arrangement, while the latter is an universal simulator for any system interacting with the given environment. Consequently, whether the objective surrogate field exists and what are its properties is determined exclusively by the environment. Thus, we are able to formulate the sufficient criterion for the environment to facilitate its surrogate, and we identify a number of environment types that satisfy it. Finally, we discuss in what sense the objective surrogate field representation can be considered classical and we explain its relation to the formation of system--environment entanglement, and the back-action exerted by the system onto environment.
\end{abstract}
		
\maketitle
	
\section{Introduction}

In recent years, we have witnessed a tremendous pace of advancement in the field of quantum technology; presently, devices that utilize quantum effects to perform useful tasks in practical circumstances seem to be an inevitable part of not so far future~\cite{Preskill_Quantum2018,Acin_NJP18}. Then, it is only natural that the focus of modern applied quantum theory shifts from idealized closed systems to more realistic {\it open systems} where the system ($S$) of interest---which can be a component of a quantum device---undergoes evolution due to application of various control protocols while experiencing the decoherence caused by the contact with its uncontrolled environment~($E$). Indeed, even the system that has been prepared, and is handled, with the utmost care is extremely unlikely to be perfectly isolated from the environment. Moreover, unlike classical systems, even weak interactions with the environment can lead to decoherence effects that fundamentally alter the properties of quantum system~\cite{Schlosshauer_book,Zurek_RMP03}. Therefore, the development of effective and accurate description of the dynamics of open quantum systems is of paramount importance.

The standard physically-motivated approach to the problem of open system dynamics is to begin with an exact two-party Hamiltonian; here, we focus on simple, albeit by no means trivial, form
\begin{align}
\label{eq:H_Q}
\hat H_{SE} = \hat H_S\otimes \hat{\mathds{1}} +\hat{V}_S\otimes\hat{V}_E +\hat{\mathds{1}}\otimes\hat H_E,
\end{align}
where $\hat H_E$, $\hat H_S$ are the free Hamiltonians of the environment and the system (with the latter incorporating any applied control scheme) and $\hat V_S$, $\hat V_E$ are the system and the environment sides of the coupling. However, in overwhelming majority of cases it is impossible to solve the dynamical problem defined by such a Hamiltonian exactly, and hence, the success relies heavily on approximation schemes---e.g., the quantum master equation method~\cite{Breuer_book_Master}---that are specific to a given system--environment arrangement (i.e., the specific choice of $\hat H_S$, $\hat H_E$, $\hat V_S\otimes\hat V_E$ and the initial state $\hat\rho_S\otimes\hat \rho_E$). The drawback is that successful schemes and techniques developed for one arrangement only rarely can be reused for treating different arrangements, even when the only modified element is the control scheme applied to the system (e.g., see~\cite{Kolodynski_PRA2018}).

An alternative approach, and the main concern of this paper, is the {\it noise representation}. Essentially, it is an attempt at assigning the involved parties with the distinct roles they play in the dynamics---the environment is viewed as the ``influencer'', or the ``driver'', and the system is the ``influencee'' or the ``driven''. In formal terms, this idea is implemented by replacing the exact description $\hat H_{SE}$ with an effective system-only Hamiltonian where the environmental operators have been superseded by an external field~\cite{Klauder_PR62,Biercuk_JoPB2011,Szankowski_JPCM2017,Degen_RMP2017,Glaser_EPJD2015,Szankowski_PRA2016,Krzywda_PRA2017,Krzywda_NJP2019,Norris_PRL2016,Norris_PRA2018,Gu_JChemPhys2019}
\begin{equation}
\label{eq:H_C}
\hat H_\xi(t) = \hat H_S + \xi(t)\hat V_S.
\end{equation}
We will refer to this field as the {\it surrogate field}.

If the surrogate field $\xi(t)$ comprises stochastic elements (as it is often the case), then $\hat H_\xi(t)$ model has to be supplemented with an {\it averaging procedure} where any quantity computed for given realization (or trajectory) of $\xi$ must be averaged over all such realizations. Assuming that only the dynamics of system-only observables are of concern, then the averaging can be incorporated into the description by adopting the following definition of the density matrix of $S$
\begin{equation}
\hat\varrho_S(t) = \overline{\hat U(t|\xi)\hat\rho_S\hat U^\dagger(t|\xi)}, \label{eq:RU}
\end{equation}
where $\hat\rho_S$ is the initial state, the unitary evolution operator conditioned by the field realization is given by a standard time-ordered exponential
\begin{equation}
\label{eq:U_cond_xi}
\hat U(t|\xi) =\mathcal{T}e^{-i\int_0^t d\tau\hat H_\xi(\tau)}= \mathcal{T}e^{-i\int_0^t d\tau [\hat H_S+\xi(\tau)\hat V_S]},
\end{equation}
and the symbol $\overline{(\ldots)}$ indicates the average over field's trajectories.

The conditions for applicability of noise representation are currently not well understood. Over the years, a few hypotheses have been posed in the literature~\cite{Audenaert_NJP08,Neder_PRB11,Zhao_PRL2011,Reinhard_PRL12,DasSarma_PRB2014,Kayser_PRA15,Hernandez_PRB18,Bethke_arXive2019,Ma_PRB2015,Chen_PRL18,Chen_NC19,Gu_JChemPhys2019}, but the definite answer has proven to be elusive. The goal of this paper is to provide a possibly complete answer to this question and to quantify the conditions under which the system--environment arrangement facilitates a valid surrogate field representation. We will not provide an in-depth discussions how one should proceed once the valid surrogate field has been found; there already exists a sizable body of literature on this particular topic that supply a number of specific examples~\cite{Klauder_PR62,Biercuk_JoPB2011,Szankowski_JPCM2017,Degen_RMP2017,Glaser_EPJD2015,Szankowski_PRA2016,Krzywda_PRA2017,Krzywda_NJP2019,Norris_PRL2016,Norris_PRA2018}. Our first and foremost concern is to find out when the surrogate field is a valid representation in the first place. We consider the representation to be valid when $\hat H_\xi$ model allows for high fidelity simulation of the actual dynamics governed by $\hat H_{SE}$ of any system-only observable; formally
\begin{equation}
\label{eq:crit}
\mathrm{tr}_E\left(e^{-i t\hat H_{SE}}\hat\rho_S\otimes\hat\rho_E\,e^{i t\hat H_{SE}}\right)
	\approx \overline{\hat U(t|\xi)\hat\rho_S\hat U^\dagger(t|\xi)},
\end{equation}
where $\hat\rho_E$ is the initial state of the environment and $\mathrm{tr}_E(\ldots)$ is the partial trace over $E$ subspace. The validity criterion we are presenting is based on the discovery of formal analogy between the average over trajectories of stochastic process and the partial trace operation. On the one hand, the average can be written as a path integral,
\begin{align}\label{eq:avg_func_int}
    &\overline{\hat U(t|\prm)\hat\rho_S\hat U^\dagger(t|\prm)}
    =\int\mathcal{D}\prm\,P[\prm]\hat U(t|\prm)\hat\rho_S\,\hat U^\dagger(t|\prm),
\end{align}
where the non-negative $P[\prm]$ is the probability distribution of the trajectories of process $\prm$. On the other hand, as we will demonstrate in Sec.~\ref{sec:objective}, the partial trace can also be cast in a form that resembles the average-like integral over trajectories
\begin{align}
    \label{eq:trace_func_int}
    &\mathrm{tr}_E\left(e^{-it\hat H_{SE}}\hat\rho_S\otimes\hat\rho_E \,e^{it\hat H_{SE}}\right)
        =\int\mathcal{D}\prm\int\mathcal{D}\bs\,Q[\prm,\bs]\,\mathscr{U}(t|\prm,\bs)\hat\rho_S,
\end{align}
where the super-operator $\mathscr{U}(t|\prm,\bs)$ (i.e., operator acting on operators) is conditioned by real-valued trajectories of a two-component ``quantum process'' $(\prm(t),\bs(t))$ and it satisfies $\mathscr{U}(t|\prm,\prm)\hat\rho_S = \hat U(t|\prm)\hat\rho_S\hat U^\dagger(t|\prm)$. The complex-valued $Q[\prm,\bs]$ plays the role analogous to the probability distribution in~\eqref{eq:avg_func_int}---essentially, it is the {\it quasi-probability distribution} of trajectories of quantum process. Then, the validity criterion specifies the sufficient conditions for the quasi-probability to be treated as a proper probability distribution, $Q[\prm,\bs]\approx \delta(\prm-\bs)P[\prm]$, and thus, for the partial trace~\eqref{eq:trace_func_int} to take on the form of the stochastic average~\eqref{eq:avg_func_int}. In Secs.~\ref{sec:joint_prob_dist} and \ref{sec:joint_quasi_dist} we will introduce an equivalent formulation of this idea that is less succinct but, ultimately, more useful in practice. Instead of functional distributions of abstract trajectories, we will make use of an alternative description where one lists the hierarchy of probabilities (or quasi-probabilities) that the trajectory passed through given sequence of values, e.g., a distribution $p_\prm^{(k)}(\prm_1t_1;\prm_2 t_2;\ldots;\prm_k t_k)$ describes the probability that trajectory of stochastic process $\prm(t)$ have value $\prm_k$ at $t_k$, followed by $\prm_{k-1}$ at $t_{k-1}$, etc.. The upside is that these joint (quasi-)probability distributions, as they are called, are standard functions and, in our case, are defined with closed analytical formulas derived from the dynamical laws of the environment (i.e., the triplet of operators $\hat V_E$, $\hat H_E$ and $\hat\rho_E$).

In the most basic terms, the surrogate field representation exchanges an exact two-party Hamiltonian for an effective evolution generator for the single party $S$. Generally speaking, the details of the representation depend on all elements of the system--environment arrangement: the initial states $\hat\rho_S$ and $\hat\rho_E$, the free Hamiltonians $\hat H_S$ and $\hat H_E$, and, of course, both the system and the environment sides of the coupling $\hat V_S$ and $\hat V_E$. It seems quite obvious that whenever the original environment is swapped for some other physical system---in the sense that any number of elements among $\hat H_E$, $\hat V_E$ and $\hat \rho_E$ are swapped for different operators---the fundamental changes to the representation should also be expected (assuming it would even still exist); after all, the field $\xi(t)$ is supposed to be a surrogate for the environment. It is much less obvious what happens to the representation when it is the system side of the arrangement that is modified, instead. This brings about the question of {\it objectivity}, or more precisely, of {\it inter-subjectivity} of surrogate field representation: for a fixed environment, how does the surrogate field depend on the {\it context} defined by the system? By context we mean here the choice of all the elements constituting the system $S$, its initial state $\hat\rho_S$, the free Hamiltonian $\hat H_S$, and the coupling operator~$\hat V_S$. Whenever any of those elements is modified, we will consider it a different context. These questions are of particular relevance for many practical applications. For example, in the field of quantum sensing~\cite{Degen_RMP2017,Szankowski_JPCM2017}, a standard approach is to employ a simple quantum system as a probe that  gathers information about the environment~$E$. Then, one attempts to utilize this information to predict the course of decoherence of a more complex systems open to~$E$. When the surrogate field representation is valid in the context of the probe, the acquired information would include the characteristics of the surrogate. If the representation happens to be context-independent, then one would be able to apply this information to simulate the evolution of an arbitrary open quantum systems. Therefore, the issue of surrogate field's objectivity is a vital one. The quasi-probability formalism~\eqref{eq:trace_func_int} we are presenting here is, by design, an ideal approach for resolving this problem. Since the quasi-probability distribution $Q[\prm,\bs]$ is wholly defined by the environment side of the $SE$ arrangement, all its properties, including the compliance with the surrogate validity criterion, are independent of the context. In essence, the decomposition~\eqref{eq:trace_func_int} is the most general implementation of the inflencer--influencee paradigm; its strength lies in the ability to identify and neatly isolate all contributions from the environment that affect the dynamics of the open system.

The system--environment coupling in~\eqref{eq:H_Q} is not of the most general form. However, historically, the noise representations have been considered for almost no other form of coupling~\cite{Klauder_PR62,Biercuk_JoPB2011,Szankowski_JPCM2017,Degen_RMP2017,Glaser_EPJD2015,Szankowski_PRA2016,Krzywda_PRA2017,Krzywda_NJP2019,Norris_PRL2016,Norris_PRA2018}, and the reason for this is not straightforward. The explanation is rather technical and referential to the findings for the coupling~\eqref{eq:H_Q}; we discuss it in Sec.~\ref{sec:mulit-component} where we investigate the prospects of a multi-component surrogate field representation for the general form of coupling.

Note that traditionally the {\it surrogate field} is referred to in the literature as the {\it classical noise}~\cite{Klauder_PR62,deSousa_TAP09,Cywinski_PRB2008,Biercuk_JoPB2011,Szankowski_QIP2015,Paz-Silva_NJP2016,Szankowski_JPCM2017}. It is then contrasted with the {\it quantum noise}~\cite{Paz-Silva_PRA2017,Beaudoin_PRA2018,Szankowski_PRA2019} which often simply means that the noise representation fails, and one has to solve the dynamics defined by the two-party Hamiltonian. However, some authors~\cite{Gardiner_book,Schoelkopf_spectrometer,Clerk_RMP10} reserve this name for the specific arrangement of a two-level system coupled with a thermal reservoir of independent quantum harmonic oscillators---the so-called spin-boson model. Here, we have chosen to abandon the traditional nomenclature because of the risk that connotations of the adjective ``classical''  might be too suggestive, and that they could provoke one to draw some far fetched conclusions about the system--environment arrangement on the basis of the name alone. For example, one might expect that there is a link between the validity of ``classical noise'' representation and ``classicality'' of the environment (which is not necessarily the case, as demonstrated in Sec.~\ref{sec:conditions}), or that the ``classical noise'' representation is incompatible with the formation of system--environment entanglement because entanglement is a ``non-classical'' type of correlation (we challenge this sentiment in Sec.~\ref{sec:entanglement}). On the other hand, the name ``surrogate field'' is not burdened by such a baggage, and it represents exactly what it advertises---the surrogate field is a surrogate for the environmental degrees of freedom. Nevertheless, some analogies between the surrogate field representation and classical theories are expected, and so, we explore this issue in Sec.~\ref{sec:loc_real}.

Further note regarding the nomenclature, the completely positive trace-preserving dynamical map established by Eq.~\eqref{eq:RU}, belongs to the class of {\it random unitary} maps~\cite{Audenaert_NJP08}. The ``prototypical'' scenario described with random unitary map, one that stems from the basic physical interpretation, occurs when the system dynamics are generated by an {\it actual} single-party Hamiltonian of form~\eqref{eq:H_C}, and where any stochasticity of the external field (as well as the averaging procedure) depicts uncertainty or ignorance of the observer~\cite{Audenaert_NJP08}.  In the terms we used here, such a scenario is described in the following way. (i)~There is no fundamental $\hat H_{SE}$ that is being replaced by $S$-only Hamiltonian $\hat H_\xi$, i.e., $\xi$ is not a surrogate for any environment but is a genuine external field, instead. (ii)~Any single instance of the system's evolution is given by unitary $\hat U(t|\xi)$ where the trajectory of $\xi$ is specified, and could be uncovered in principle, but is unknown to the observer. (iii)~Because of this uncertainty, any expectation value predicted by the observer has to be averaged over all possibilities. Based on this example, some authors~\cite{Audenaert_NJP08,Kayser_PRA15} choose to classify evolution maps as {\it classical} when the map can be written in random unitary form, and as {\it truly quantum} or {\it non-classical} otherwise. The discussion on the relationship between surrogate field representation and the formation of system--environment entanglement presented in Sec.~\ref{sec:entanglement} provides an argument that such a classification scheme could be enriched with additional nuance. Overall, surrogate field representations showcased in Sec.~\ref{sec:conditions} also provide a number of non trivial examples that should be useful for developing intuitions regarding the underlying physics of random unitary map theory.

\section{Objective surrogate field representation}\label{sec:objective}

\subsection{Joint probability distributions}\label{sec:joint_prob_dist}

We begin by examining the structure of system state $\hat\rho_S(t)$ resulting from the simulation with stochastic Hamiltonian~$\hat H_\xi(t) = \hat H_S+\prm(t)\hat V_S$. Switching to the interaction picture $\hat\rho_S^I(t) = e^{it\hat H_S}\hat\rho_S(t)e^{-it\hat H_S}$ and expanding the time-ordered exponentials in $\hat U(t|\prm)$ into series, we obtain the following form of the density matrix
\begin{align}
\nonumber
\hat\rho_S^I(t) &=e^{it\hat H_S}\overline{\hat U(t|\prm)\hat\rho_S\hat U^\dagger(t|\prm)}e^{-it\hat H_S}\\
\nonumber
&=\sum_{k=0}^\infty(-i)^k\int_0^tdt_1\int_0^{t_1}dt_2\ldots\int_0^{t_{k-1}}\!\!\!dt_k\,
	\overline{\prm(t_1)\ldots\prm(t_k)}
	\left(\prod_{l=1}^k\mathscr{V}_S(t_l)\right)\hat\rho_S\\
\label{eq:rhoS_stochastic}
&=\sum_{k=0}^\infty(-i)^k\int_0^tdt_1\int_0^{t_1}dt_2\ldots\int_0^{t_{k-1}}\!\!\!dt_k
	\sum_{\boldPrm\in\Omega_\prm^{\times k}}p_\prm^{(k)}(\boldPrm\boldT)
	\left(\prod_{l=1}^k \prm_l \mathscr{V}_S(t_l)\right)\hat\rho_S,
\end{align}
where the super-operators acting on the initial state are defined as
\begin{align}
\label{eq:V_superop}
\mathscr{V}_S( t)\hat A = \hat V_S(t)\hat A - \hat A \hat V_S(t),
\end{align}
with $\hat V_S(t) = e^{it\hat H_S}\hat V_Se^{-it\hat H_S}$, and the symbol $\prod_{l=1}^k\mathscr{V}_S(t_l)$ applied to super-operators is understood as an ordered composition $\mathscr{V}_S(t_1)\mathscr{V}_S(t_2) \ldots \mathscr{V}_S(t_k)$. The influence of the noise on the course of the evolution is quantified by the family of {\it joint probability distributions}
\begin{align}
p_\prm^{(k)}(\boldPrm\boldT) = p_\prm^{(k)}(\prm_1t_1;\prm_2t_2;\ldots;\prm_kt_k),
\end{align}
which establish the probability of the process~$\xi(t)$ having the value $\xi_k$ at the initial time $t_k$, followed by $\xi_{k-1}$ at $t_{k-1}$, ..., and terminating with $\xi_1$ at $t_1$ (assuming $t_1>t_2>\ldots>t_k$); the range of values available to $\prm(t)$ is constraint by the set $\Omega_\xi$. The family of probability distributions $\{ p_\xi^{(k)} \}_{k=1}^\infty$ defines $\xi(t)$ completely, and thus, functions $p_\xi^{(k)}$ cannot be arbitrary as they have to satisfy the following two conditions~\cite{vanKampen_book}:

(i) Since each $p^{(k)}_\xi$ is a probability distribution, it has to be {\it non-negative}
\begin{equation}
p^{(k)}_\xi(\boldPrm\boldT)\geqslant 0,
\end{equation}
for all $\boldPrm\in\Omega_\prm^{\times k}$ and $t_1>t_2>\ldots>t_k$.

(ii) The joint probabilities belonging to one family are related through Chapman-Kolmogorov {\it consistency criterion}
\begin{align}
\label{eq:CK_consistency}
&\sum_{\xi_l\in\Omega_\xi} p^{(k)}_\xi(\boldPrm\boldT)
	=p^{(k-1)}_\xi(\ldots;\xi_{l-1}t_{l-1};\xi_{l+1}t_{l+1};\ldots),
\end{align}
for $t_1>t_2>\ldots>t_k$, and
\begin{align}
\sum_{\xi_1\in\Omega_\xi}\! p^{(1)}_\xi(\xi_1t_1) = 1.
\end{align}
Conversely, any set of functions that satisfy both of the above conditions defines some stochastic process. This fact will be the linchpin of our search for surrogate field representation.

\subsection{Joint quasi-probability distributions}\label{sec:joint_quasi_dist}

The next step is to express the reduced system state evolved under the two-party Hamiltonian $\hat H_{SE}$ in a form that will most directly compare with the previously obtained expression~\eqref{eq:rhoS_stochastic}. We demonstrate in the Methods Sec.~\ref{sec:rhoS_quantum} that the interaction picture of the reduced density matrix can be written in the following way
\begin{align}
\nonumber
\hat\rho_S^I(t)&= e^{it\hat H_S}\,\mathrm{tr}_E\left(
	e^{-it\hat H_{SE}}	\hat\rho_S\otimes\hat\rho_E\,e^{it\hat H_{SE}}\right)e^{-it\hat H_S}\\
\label{eq:rhoS_quantum}
&=\sum_{k=0}^\infty (-i)^k \int_0^t dt_1\int_0^{t_1}\!\!dt_2\ldots\int_0^{t_{k-1}}\!\!\!dt_k
	\sum_{\boldPrm,\boldBs\in\Omega_{\hat V}^{\times k}}\delta_{\prm_1,\bs_1}\,
	\objQ{k}(\boldPrm\boldBs\boldT)
	\left(\prod_{l=1}^k\mathscr{W}_S(\prm_l\bs_lt_l)\right)\hat\rho_S.
\end{align}
Here, the set $\Omega_{\hat V}$ is the spectrum of the environment-side coupling operator $\hat V_E$ and it contains all of its unique eigenvalues
\begin{align}
\hat V_E = \sum_n v_n |n\rangle\langle n| = \sum_{\prm\in\Omega_{\hat V}}\prm \sum_{n:\prm = v_n}|n\rangle\langle n|.
\end{align}
The action of super-operators $\mathscr{W}_S(\prm\bs t)$ is defined by
\begin{align}
\label{eq:W_superop}
\mathscr{W}_S(\prm\bs t)\hat A &=\frac{1}{2}(\prm+\bs)\left(\hat V_S(t)\hat A - \hat A\hat V_S(t)\right)+
	\frac{1}{2}(\prm-\bs)\left(\hat V_S(t)\hat A + \hat A\hat V_S(t)\right),
\end{align}
and they encapsulate the explicitly context-dependent contribution to the evolution. Finally, the family of functions $\{\objQ{k}\}_{k=1}^\infty$, which we will call the {\it joint quasi-probability distributions}, are given by
\begin{align}
\nonumber
\objQ{k}(\boldPrm\boldBs\boldT) & =\objQ{k}(\prm_1\bs_1t_1;\prm_2\bs_2t_2;\ldots;\prm_k\bs_k t_k)\\
\label{eq:q_def}
&=\Bigg(\prod_{l=1}^k\, \sum_{\substack{n_l:\\ \prm_l=v_{n_l}}}\sum_{\substack{m_l:\\ \bs_l=v_{m_l}}}\Bigg)\delta_{n_1,m_1}
	\langle n_k|\hat\rho_E(t_k)|m_k\rangle\left(\prod_{l=1}^{k-1}T_{t_{l}-t_{l+1}}(n_{l} m_{l}|n_{l+1}m_{l+1})\!\right),
\end{align}
where $\hat\rho_E(t) = e^{-i t\hat H_E}\hat\rho_E e^{i t\hat H_E}$ and the {\it propagator}
\begin{align}
\label{eq:propagator_def}
T_t(nm|n'm') &= \mathrm{tr}_E\left( |m\rangle\langle n|\, e^{-it\hat H_E}|n'\rangle\langle m'|e^{it\hat H_E} \right),
\end{align}
constitute the fundamental context-independent building blocks of the whole structure.

The key feature of Eq.~\eqref{eq:rhoS_quantum} is the analogy between families of quasi-probabilities $\{ \objQ{k}\}_{k=1}^\infty$ and proper probabilities $\{ p^{(k)}_\xi \}_{k=1}^\infty$ which goes beyond simplistic formal resemblance of the corresponding formulas. Indeed, in the Methods Sec.~\ref{sec:quasi-CK_crit} we verify that, just like joint probabilities, functions $\objQ{k}$ satisfy the Chapman-Kolmogorov consistency criterion
\begin{align}
\label{eq:q_CK}
&\sum_{\prm_l,\bs_l\in\Omega_{\hat V}} \objQ{k}(\boldPrm\boldBs\boldT)
	=\objQ{k-1}(\ldots;\prm_{l-1}\bs_{l-1}t_{l-1};\prm_{l+1}\bs_{l+1}t_{l+1};\ldots).
\end{align}
However, as we will see below, quasi-probabilities do not necessarily satisfy the condition of non-negativity. Therefore, in general, a given family $\{\objQ{k}\}_{k=1}^\infty$ does not properly define a stochastic process.

\subsection{Structure of joint quasi-probability distributions}\label{sec:structure}

When examined as a diagram, $\objQ{k}$ can be viewed as a superposition of a selection of {\it propagator chains} (or simply {\it chains}) where propagators $T_t(nm|n'm')$ play the role of chain links and the connections between consecutive links are established through {\it projectors}~$|n\rangle\langle n|$ or {\it coherences}~$|n\rangle\langle m|$ ($n\neq m$) that enforce the matching of indices
\begin{align}
&\ldots\overbrace{T_{t-t'}(nm\underbrace{|n'm')T_{t'-t''}(n'm'|}_{\mathclap{\text{connection through $|n'\rangle\!\langle m'|$}}}n''m'')}^{\text{a two-link segment of the chain}}\ldots.
\end{align}
Each chain begins with a special link in a form of density matrix element and ends with a propagator that carries a disconnected projector
\begin{align}
\label{eq:beg_end}
&T_{t_1-t_2}\overbrace{(n_1n_1|}^{\mathclap{\text{disconnected projector $|n_1\rangle\!\langle n_1|$}}}n_2m_2)T_{t_2-t_3}(n_2m_2|n_3m_3)\ldots
	 T_{t_{k-1}-t_k}(n_{k-1}m_{k-1}|n_km_k)
	\underbrace{\langle n_k|\hat\rho_E(t_k)|m_k\rangle}_{\mathclap{\text{initial link}}}.
\end{align}
In general, propagators are complex functions
\begin{align}
\label{eq:Schroedinger_propagators}
T_t(nm|n'm') = \langle n|e^{-it\hat H_E}|n'\rangle \langle m|e^{-it\hat H_E}|m'\rangle^*,
\end{align}
and hence, the chains cannot be assigned with a definite sign (in particular, they are not necessary non-negative). However, among all the chains constituting a given quasi-probability distribution we can distinguish a class composed entirely of propagators connected through projectors---the {\it projector-connected chains}---such that each link is of the form
\begin{align}
T_t(nn|mm)=\big|\langle n|e^{-it \hat H_E}|m\rangle\big|^2\geqslant 0,
\end{align}
including the initial link $\langle n_k|\hat\rho_E(t_k)|n_k\rangle\geqslant 0$. Consequently, the sum of all such chains is {\it non-negative}
\begin{align}
\label{eq:P_SE}
\objP{k}(\boldPrm\boldT)\equiv&
	\Bigg(\prod_{l=1}^k\,\sum_{\substack{n_l: \prm_l=v_{n_l}}}\Bigg)\langle n_k|\hat\rho_E(t_k)|n_k\rangle
	\left(\prod_{l=1}^{k-1}T_{t_l-t_{l+1}}(n_ln_l|n_{l+1}n_{l+1})\right)
	\geqslant 0.
\end{align}
Although $\objP{k}(\boldPrm\boldT)$'s are a multivariate probability distributions (they are non-negative and normalized), they do not form a proper family of joint probabilities which would allow to interpret the series of random variables $\prm_1,\prm_2,\ldots,\prm_k$ as a sample of stochastic process $\prm(t)$. Indeed, the remainder $\Delta\objQ{k}(\boldPrm\boldBs\boldT)$ that consists of all the chains with, at least, one connection through coherence $|n\rangle\langle m|$ ($n\neq m$), as defined by the decomposition
\begin{align}
\label{eq:Q_decomp}
\objQ{k}(\boldPrm\boldBs\boldT) = \delta_{\boldPrm,\boldBs}\,\objP{k}(\boldPrm\boldT)+\Delta \objQ{k}(\boldPrm\boldBs\boldT),
\end{align}
has to be also taken into account for the consistency criterion to be satisfied. On the other hand, it is the contribution from those {\it coherence-connected propagator chains} that hinders the compliance with the non-negativity criterion as $\Delta \objQ{k}$, in contrast to $\objP{k}$, cannot be guaranteed to have a definite sign. From an oversimplified point of view, the ``quantumness'' of a two-component quantum process $(\prm(t),\bs(t))$ manifests itself through coherence-connected chains as their composition $\Delta \objQ{k}$ is a proper quantum superposition where the amplitudes for alternative outcomes have the ability to interfere with each other.

\subsection{From quantum process to surrogate field}\label{sec:crit}
Suppose now that $\Delta \objQ{k}$'s are negligible and the quasi-probabilities become compliant with the non-negativity criterion $\objQ{k}(\boldPrm\boldBs\mathbf{t})\approx \delta_{\boldBs,\boldPrm}\,\objP{k}(\boldPrm\mathbf{t})\geqslant 0$. In such a case, the remaining probability distributions $\objP{k}$ satisfy the Chapman-Kolmogorov consistency criterion by themselves
\begin{align}
\nonumber
\sum_{\prm_l\in\Omega_{\hat V}}\objP{k}(\boldPrm\boldT) &\approx \sum_{\prm_l,\bs_l\in\Omega_{\hat V}}\objQ{k}(\boldPrm\boldBs\boldT)
	= \objQ{k-1}(\ldots;\prm_{l-1}\bs_{l-1}t_{l-1};\prm_{l+1}\bs_{l+1}t_{l+1};\ldots)\\
&\approx \objP{k-1}(\ldots;\prm_{l-1}t_{l-1};\prm_{l+1}t_{l+1};\ldots),
\end{align}
and, as a result, they form the joint probability distribution family $\{\objP{k}\}_{k=1}^\infty$ that defines stochastic process $\prm(t)$. More importantly, this stochastic process is actually a surrogate field that simulates the evolution of the reduced state of $S$ via the model Hamiltonian $\hat H_\prm(t)=\hat H_S+\prm(t)\hat V_S$. Indeed, since $\mathscr{W}_S(\prm\prm t) = \prm\mathscr{V}_S(t)$ [compare Eqs.~\eqref{eq:V_superop} and \eqref{eq:W_superop}], the reduced density matrix becomes
\begin{align}
\nonumber
\hat\rho_S^I(t) &= e^{it\hat H_S}\,\mathrm{tr}_E\left(
	e^{-it\hat H_{SE}}	\hat\rho_S\otimes\hat\rho_E\,e^{it\hat H_{SE}}\right)e^{-it\hat H_S}\\
\nonumber
&\approx\sum_{k=0}^\infty(-i)^k \int_0^tdt_1\ldots\int_0^{t_{k-1}}\!\!\!dt_k
	\sum_{\boldPrm,\boldBs\in\Omega_{\hat V}^{\times k}}\!\! \delta_{\boldPrm,\boldBs}\,\objP{k}(\boldPrm\boldT)
	\left(\prod_{l=1}^k\mathscr{W}_S(\prm_l\bs_lt_l)\right)\hat\rho_S\\
\nonumber
&=\sum_{k=0}^\infty(-i)^k \int_0^tdt_1\ldots\int_0^{t_{k-1}}\!\!\!dt_k
	\sum_{\boldPrm\in\Omega_{\hat V}^{\times k}}
	\objP{k}(\boldPrm\boldT)\left(\prod_{l=1}^k\prm_l\mathscr{V}_S(t_l)\right)\hat\rho_S\\
&=e^{it\hat H_S}\overline{\hat U(t|\prm)\hat\rho_S\hat U^\dagger(t|\prm)}e^{-it\hat H_S},
\end{align}
which holds true in any context (i.e., for any choice of $\hat H_S$, $\hat V_S$ and $\hat\rho_S$). In fact, the joint quasi-probabilities can be considered context-independent: each $\objQ{k}$ is defined completely and exclusively by the environment side of $SE$ arrangement (hence, the index $E$). Therefore, whether $\Delta\objQ{k}\approx 0$ and $\objQ{k}$'s form the stochastic process-defining family of joint probabilities is determined only by the properties of the environment, and if they do, then the evolution of any system coupled to $E$ through operator $\hat V_E$ is simulable with the same surrogate field. In other words, the surrogate created in this way is inter-subjective in all contexts.

We will now summarize the above deliberations with formally stated sufficient criterion for validity of the objective surrogate field representation; this criterion can be considered as the main result of the paper. 
\begin{theorem}[Objective surrogate field representation]\label{crt:obj}
For the environment $E$ (defined by $\hat V_E$, $\hat H_E$, and $\hat\rho_E$) to facilitate the {\it objective surrogate field representation} $\prm(t)$---the representation that is inter-subjective in all contexts---it is sufficient that, for each member of the family of joint quasi-probability distributions $\{ \objQ{k} \}_{k=1}^\infty$, the superposition of coherence-connected propagator chains $\Delta \objQ{k}$ is negligible so that
\begin{equation*}
\objQ{k}(\boldPrm\boldBs\boldT)\approx \delta_{\boldPrm,\boldBs}\,\objP{k}(\boldPrm\boldT).
\end{equation*}
Then, the stochastic process $\prm(t)$ is defined by the family of joint probability distributions~$\{\objP{k}\}_{k=1}^\infty$.
\end{theorem}

When the surrogate representation is valid, and the environmental Hamiltonian $\hat H_E$, the initial state $\hat\rho_E$, and the eigensystem of the coupling $\{|n\rangle\}_n$, $\Omega_{\hat V}$ are known, then, in principle, the following algorithm allows to instantiate trajectories of surrogate field $\xi(t)$: (i)~Choose an arbitrary time grid $\mathbf{t}_\mathrm{grd} = (t_1,\ldots,t_k)$ and $t_1>\ldots>t_k$. (ii)~Calculate the joint probability distribution $\objP{k}(\boldPrm\mathbf{t}_\mathrm{grd})$ according to \eqref{eq:P_SE} for all values of $\boldPrm\in \Omega_{\hat V}^{\times k}$. Although straightforward, this is the most difficult and resource intensive step. (iii)~Draw at random from previously obtained distribution the sequence $\boldPrm_\mathrm{smp} = (\prm_1,\ldots,\prm_k)$; such a sequence is a sample trajectory of the process spanned on grid $\mathbf{t}_\mathrm{grd}$. This concludes the procedure.

Once the probability distribution has been successfully calculated in the second step of the above procedure, the last step can be repeated any number of times at relatively low cost. The resultant ensemble of sample trajectories---provided the time grid is fine enough and the number of samples is sufficiently large---can be used to carry out the averaging procedure of any quantity. This includes not only the expectation values of system-only observables, but also quantities that characterize the process itself, like its moments or cumulants. In Methods Sec.~\ref{sec:tutorial}, we give a basic overview how this procedure is utilized for finding the system dynamics.

\section{Examples of environments that facilitate objective surrogate field}\label{sec:conditions}

\subsection{Quasi-static coupling}\label{sec:quasi-static}

Assume that the environmental free Hamiltonian and the coupling operator commute
\begin{align}
[\hat H_E,\hat V_E]=0.
\end{align}
Then, the eigenstates of the coupling operator $|n\rangle$ are, simultaneously, eigenstates of the Hamiltonian $\hat H_E|n\rangle = \epsilon_n|n\rangle$. It follows that, within each propagator, the evolution operators preserve the orthogonality between projectors $|n\rangle\langle n|$ and coherences $|m\rangle\langle  m'|$ ($m\neq m'$)
\begin{align}
T_t(nn|mm') & = \mathrm{tr}_E\big(|n\rangle\langle n|e^{-i t\hat H_E}|m\rangle\langle m'|e^{i t\hat H_E}\big)
	=e^{i t(\epsilon_{m'}-\epsilon_{m})}\mathrm{tr}_E\big(|n\rangle\langle n||m\rangle\langle m'|\big) = 0.
\end{align}
Therefore, all coherence-connected chains vanish because each one of those chains contains at least one instance of propagator linking a coherence and a projector [see Eq.~\eqref{eq:beg_end}]. In such a case, any superposition of those chains, including $\Delta\objQ{k}$'s, vanishes as well. On the other hand, the projector-connected chains (and their combinations) are preserved, and hence, the joint quasi-probabilities become a proper probability distributions $\objQ{k}(\boldPrm\boldBs\mathbf{t}) = \delta_{\boldPrm,\boldBs}\objP{k}(\boldPrm\mathbf{t})$ that read
\begin{align}
\objP{k}(\boldPrm\mathbf{t}) =
	\left(\prod_{l=2}^k\delta_{\prm_1,\prm_l}\right) \!\sum_{\substack{n: \prm_1=v_n}}\!\langle n|\hat\rho_E|n\rangle.
\end{align}
The resultant surrogate field $\prm$ is of the {\it quasi-static} noise type---a stochastic process that is time-independent (essentially, a random variable). The process is governed by the probability distribution $p(\prm) = \sum_{n:\prm=v_n}\langle n|\hat\rho_E|n\rangle$ given by the initial state of $E$ and the range of values that coincide with the spectrum of coupling operator $\Omega_\prm=\Omega_{\hat V}$.

\subsection{Open environment}\label{sec:open}

Suppose that the environmental degrees of freedom can be further separated into two subspaces: one that is in direct contact with the system (let us still label it as $E$), and the other part ($D$) that is decoupled from the system but interacts with~$E$
\begin{align}
\label{eq:driven_Hamiltonian}
\hat H &= \hat H_S\otimes\hat{\mathds{1}}\otimes\hat{\mathds{1}}
	+\hat V_S\otimes\hat V_E\otimes\hat{\mathds{1}}
	+\hat{\mathds{1}}\otimes\hat H_E\otimes\hat{\mathds{1}}
	+\hat{\mathds{1}}\otimes\hat V_{ED}+\hat{\mathds{1}}\otimes\hat{\mathds{1}}\otimes\hat H_D.
\end{align}
Essentially, $D$ is an environment of $E$ but not of $S$.

In appendix~\ref{sec:Q_4_driven} we show that the joint quasi-probability distributions resultant from this form of environmental dynamics are given by an effective average over $D$ degree of freedom (a partial trace over $D$):
\begin{align}
\nonumber
\objQ{k}(\boldPrm\boldBs\mathbf{t}) &=\sum_{\substack{n_1: \prm_1=v_{n_1}}}
	\Bigg(\prod_{l=2}^k\,\sum_{\substack{n_l: \prm_l=v_{n_l}  }}\,\sum_{\substack{m_l: \bs_l=v_{m_l}  }}\Bigg)
		\langle n_1|\mathrm{tr}_D\Bigg[
		\Bigg(\prod_{l=2}^{k}\hat U_{ED}(t_{l-1}-t_{l})|n_{l}\rangle\langle n_{l}|\Bigg)\\
\label{eq:driven_Q}
&\phantom{=\times\langle n|}
	\times\hat U_{ED}(t_k)\hat\rho_E\otimes\hat\rho_D\hat U^\dagger_{ED}(t_k)
	\Bigg(\prod_{l=k}^2 |m_l\rangle\langle m_l|\hat U^\dagger_{ED}(t_{l-1}-t_l)\Bigg)\Bigg]|n_1\rangle,
\end{align}
where the symbol $\prod_{l=l_b}^{l_e}\hat A(l)$ applied to operators is to be read as an ordered composition: $\hat A(l_b)\hat A(l_b+1)\ldots \hat A(l_e)$ for $l_b < l_e$, or $\hat A(l_b)\hat A(l_b-1)\ldots \hat A(l_e)$ for $l_b > l_e$, and the unitary evolution operator
\begin{align}
\label{eq:ED_unitary}
\hat U_{ED}(t) = e^{-i t( \hat H_E + \hat V_{ED} + \hat H_D)},
\end{align}
operates in $ED$ subspace while the projectors $|n_l\rangle\langle n_l |$ onto eigenstates of $\hat V_E$ act only in $E$ subspace.

Assume the initial state $\hat\rho_D$ and the relation between coupling $\hat V_{ED}$ and the free Hamiltonians are such that we can invoke the {\it Born approximation} \cite{Breuer_book_Born}
\begin{align}
\label{eq:Born}
\hat U_{ED}(t)\hat A\otimes \hat\rho_D \hat U^\dagger_{ED}(t) \approx \hat{\tilde{A}}(t)\otimes \hat\rho_D.
\end{align}
In addition, in order to parametrize the undergoing dynamical process only in the terms of environment part that couples directly to the system, assume the {\it Markov} and {\it secular} approximations \cite{Breuer_book_Secular_Markov} that specify the form of the dynamical map acting on $E$
\begin{align}
\label{eq:Markov}
\hat U_{ED}(t-t')\hat A\otimes\hat\rho_D\hat U_{ED}^\dagger(t-t') \approx \left(\dMap{t,t'}\hat A\right)\otimes\hat\rho_D.
\end{align}
Here, superoperators $\dMap{t,t'}$ satisfy the composition rule
\begin{align}
\label{eq:semi-group}
\dMap{t,t'}\dMap{t',t''} = \dMap{t,t''},\ \text{for $t>t'>t''$},
\end{align}
and are generated by non-hermitian super-operator $\mathscr{L}_E(\tau)$---the so-called Lindbladian---that acts in the $E$-operators subspace only
\begin{align}
\dMap{t,t'} = \mathcal{T}e^{\int_{t'}^t \mathscr{L}_E(\tau)d\tau},\ \text{for $t>t'$.}
\end{align}
In the terms of open system theory, it is this full suite of approximations that lead to a quantum master equation for the evolution of reduced state of a system open to $D$ $\hat R_E(t) = \mathrm{tr}_D(\hat U_{ED}(t)\hat\rho_E\otimes\hat\rho_D\hat U^\dagger_{ED}(t))$,
\begin{align}
\label{eq:master_eq}
\frac{\partial}{\partial t} \hat R_E(t) = \mathscr{L}_E(t)\hat R_E(t).
\end{align}
In our case, the secular Born--Markov approximation leads to quasi-probability distributions in a standard form of propagator chain superpositions \eqref{eq:q_def} but with propagator links \eqref{eq:propagator_def} modified according to
\begin{align}
T_{t-t'}(n m|n' m') =\mathrm{tr}_E\big( |m\rangle\langle n| \dMap{t,t'}|n'\rangle\langle m'|\big),
\end{align}
and the analogous modification to the initial link where $\langle n_k|e^{-it_k\hat H_E}\hat\rho_E e^{it_k\hat H_E}|m_k\rangle$ is replaced with $\langle n_k|\dMap{t_k,0}\hat\rho_E|m_k\rangle$. Note that the composition rule~\eqref{eq:semi-group} is crucial, as it is required for quasi-probability distributions to satisfy the consistency criterion.

The fact that dynamical map $\dMap{t,t'}$ is not unitary (Lindbladian is non-Hermitian in general), opens new possibilities for breaking the coherence-connected propagator chains. One way to achieve such an effect, is for the evolution super-operator to satisfy
\begin{subequations}
\label{eq:driving}
\begin{align}
\dMap{t,t'}|n\rangle\langle n| &= \sum_{m} u_{m,n}(t,t')|m\rangle\langle m|,\\
\dMap{t,t'}|n\rangle\langle n'| &= \sum_{m\neq m'} w_{mm',nn'}(t,t') |m\rangle\langle m'|\quad(n\neq n').
\end{align}
\end{subequations}
That is, the super-operator maps projectors onto combination of projectors and coherences onto combination of coherences, thus, preserving their mutual orthogonality. When this is the case, then all coherence-connected chains constituting $\Delta\objQ{k}$ vanish because each one of them contains at least one instance of propagator of form $T_t(n_ln_l|n_{l+1}m_{l+1})\propto \delta_{n_{l+1},m_{l+1}} = 0$ ($n_{l+1}\neq m_{l+1}$); note the similarity to quasi-static coupling case from Sec.~\ref{sec:quasi-static}. Moreover, the remaining combinations of projector-connected chains $\objP{k}$ are guaranteed to be non-negative because $\dMap{t,t'}$ is a trace-preserving and completely positive map so that $u_{m,n}(t,t')\geqslant 0$ and $\sum_{m}u_{m,n}(t,t')=1$, for all $t>t'$ and $n,m$. Therefore, when environment dynamics have the property \eqref{eq:driving}, $\{\objP{k}\}_{k=1}^\infty$ is a family of proper joint probability distributions and they define a surrogate field.

The following simple example showcases how this type of environmental dynamics supports an objective surrogate field representation. Let $E$ be a two-level system that is driven by time-independent Lindbladian defined by $\mathscr{L}_E\hat A = -(\gamma/2)[\hat \sigma_x, [\hat\sigma_x,\hat A]]$ and the coupling operator is $\hat V_E =(|{+}\rangle\langle{+}|-|{-}\rangle\langle{-}|)/2 = \hat\sigma_z/2$. Then, the coupling has two eigenvalues $v_\pm=\pm 1/2$ corresponding to $\left|{\pm} \right\rangle$ eigenstates. It is a matter of straightforward algebra to verify that conditions \eqref{eq:driving} are satisfied here. The resultant probability distributions are given by
\begin{align}
\label{eq:RTN}
\objP{k}(\boldPrm\boldT)
	&= \langle \mathrm{sign}(\prm_k)|e^{t_k\mathscr{L}_E}\hat\rho_E|\mathrm{sign}(\prm_k)\rangle
	\prod_{l=1}^{k-1}\left(\frac{1+\mathrm{sign}(\prm_l)\mathrm{sign}(\prm_{l+1}) e^{-2\gamma(t_l-t_{l+1})}}{2}\right),
\end{align}
with process spectrum $\Omega_\xi = \{ {+}1/2, {-}1/2\}$. We recognize that this family of probability distributions describe a well known random telegraph noise \cite{vanKampen_book}---a stochastic process that switches between two values, $\prm= \pm 1/2$ in this case, at the rate $\gamma$.

\subsection{Environment of least action}\label{sec:action}

In the last example, we will require that the spectrum of the coupling is dense so that the sums in \eqref{eq:q_def} can be replaced with integration 
\begin{align}
\sum_{\substack{n_l:\prm_l=v_{n_l}}}\sum_{\substack{m_l:\bs_l=v_{m_l}  }} \rightarrow \int_{\Gamma_{\prm_l}}\!\!\! dx_l\int_{\Gamma_{\bs_l}}\!\!\! dy_l,
\end{align}
where the intervals $\Gamma_{\prm_l}$ are the degenerate subspaces corresponding to eigenvalues $\prm_l$ and $\Gamma_\infty$ is the whole configuration space. Using this representation we rewrite \eqref{eq:q_def} into a form that will be better suited for our current purposes
\begin{align}
\nonumber
\objQ{k}(\boldPrm\boldBs\mathbf{t}) &=	\int_{\Gamma_{\prm_1}}\!\!\! dx_1 dy_1 \delta(x_1-y_1)	
	\int_{\Gamma_\infty}\!\!\! dx_{0}\, dy_{0}\langle x_0|\hat\rho_E|y_0\rangle\\
&\phantom{=}
	\times K(x_1,x_0,\mathbf{t}|\Gamma_{\prm_2}\ldots \Gamma_{\prm_{k}})
	K(y_1,y_0,\mathbf{t}|\Gamma_{\bs_2}\ldots \Gamma_{\bs_{k}})^*,
\end{align}
where we have defined the {\it Schr\"{o}dinger chains}
\begin{align}
&K(x_1,x_0,\mathbf{t}|\Gamma_{\prm_2} \ldots \Gamma_{\prm_{k}})=\left(\prod_{l=2}^{k}\int_{\Gamma_{\prm_l}}\!\! dx_l\right)
	\left(\prod_{l=1}^{k-1}\langle x_{l}|e^{-i(t_{l}-t_{l+1})\hat H_E}|x_{l+1}\rangle\right)
	\langle x_k|e^{-i t_k \hat H_E}|x_0\rangle,
\end{align}
which are simply an alternative to the propagator chain description.

Assume the environment is such that the least action principle approximation is applicable to the Feynman path integral representation of its Schr\"{o}dinger propagators \cite{Feynman_Thesis}
\begin{align}
&\langle x_e |e^{-i (t_e-t_b)\hat H_E}|x_b\rangle
= \int_{x(t_b) = x_b}^{x(t_e) = x_e}\!\!\!\mathcal{D}x(t)\, e^{i S[x(t)]}\propto e^{i S_\mathrm{cl}(x_et_e;x_bt_b)},
\end{align}
where $S_\mathrm{cl}(x_e t_e;x_b t_b)$ is the environment action \cite{Landau_Mechanics_Action} associated with the {\it classical} trajectory of coordinate $x$---i.e., the trajectory $x(t)$ that satisfy the corresponding Euler-Lagrange equation~\cite{Landau_Mechanics_Action}---that begins at point $x_b$ at initial time $t_b$, and ends at point $x_e$ at time~$t_e$. The approximation is justified using the stationary phase method: When the action $S[x(t)]$ is large (e.g., like for massive macroscopic systems), the destructive interference between rapidly oscillating phase factors ${\exp}\{ i S[x(t)]\}$ suppresses the integration over almost all trajectories, except for the immediate vicinity of the stationary point (or rather, the stationary trajectory) of action. The least action principle of classical mechanics asserts that the trajectory which satisfies the classical equation of motion is such a stationary point (and {\it vice versa}). Therefore, the only significant contribution to the integral comes from the neighborhood of $S_\mathrm{cl}$ where the phase slows down and the interference is constructive.

First, we will consider one of the intermediate segments along the Schr\"{o}dinger chain,
\begin{align}
\nonumber
&\int_{\Gamma_{\prm_l}}\!\!\! dx_l \langle x_{l-1}|e^{-i(t_{l-1}-t_l)\hat H_E}|x_l\rangle\langle x_l|e^{-i(t_l-t_{l+1})\hat H_E}|x_{l+1}\rangle\\
&\phantom{\int_\Gamma dx}
	\propto\int_{\Gamma_{\prm_l}}\!\!\! dx_l\, e^{i S_\mathrm{cl}(x_{l-1}t_{l-1};x_lt_l)+i S_\mathrm{cl}(x_lt_l;x_{l+1}t_{l+1})},
\end{align}
where, for now, we will treat the time arguments $t_{l-1}>t_l>t_{l+1}$ and the end points $x_{l+1},x_{l-1}$ as fixed values. Since the action is large, according to stationary phase method, the integral ``stitching'' the propagators will vanish due to destructive interference, unless the degenerate subspace $\Gamma_{\prm_l}$ contains a stationary point of the phase. To determine if $x_l$ is such a point we have to check the derivative of the phase,
\begin{align}
&\frac{\partial [S_\mathrm{cl}(x_{l}t_{l};x_{l+1}t_{l+1})+S_\mathrm{cl}(x_{l-1}t_{l-1};x_lt_l)  ]}{\partial x_l} = p_e-p_b.
\end{align}
Here, we have utilized the theorem from classical theory that the derivative of the action in respect to the end/beginning point of the trajectory equals the momentum/minus momentum at the corresponding time~\cite{Landau_Mechanics_Action_as_Func}, and so, $p_e$ is the momentum at the end of trajectory from $x_{l+1}$ to $x_{l}$, and $p_b$ is the initial momentum at the beginning of trajectory from $x_{l}$ to $x_{l-1}$. In general, $p_b\neq p_e$ and $x_l$ is {\it not} a stationary point. Indeed, if we set the initial momentum at $t_l$ to $p_e$, then the coordinate would propagate, in accordance with Euler-Lagrange equation, from $x_l$ to a certain point $\tilde{x}_{l-1}$ that is different than the expected end point $x_{l-1}$. In order to make the end point match the desired $x_{l-1}$, the initial momentum has to be adjusted, which can be visualized as an application of impulse force that causes the discontinuity in momentum. However, there is one instance when such an intervention is not necessary:  $x_l$ is the stationary point (i.e., $p_e = p_b$) when it happens to lie on the classical trajectory from $x_{l+1}$ {\it directly} to $x_{l-1}$. 

We can now apply the above reasoning to the Schr\"{o}dinger chain $K(x_1,x_k,\mathbf{t}|\Gamma_{\prm_2}\ldots \Gamma_{\prm_{k}})$ as a whole. For given $\mathbf{t}$ and the end points of the trajectory $x_1$, $x_0$, the interference effects restrict the choice of $\Gamma_{\prm_2},\ldots,\Gamma_{\prm_{k}}$ to only one sequence where each interval overlaps with the classical trajectory from $x_0$ to $x_1$. Since each $\Gamma_{\prm_l}$ corresponds to eigenvalue $\prm_l$, the choice of arguments $\boldPrm$ for which $\objQ{k}(\boldPrm\boldBs\mathbf{t})\neq 0$, is identically restricted. For the same reasons, but applied to the other Schr\"{o}dinger chain $K(y_1,y_0,\mathbf{t}|\Gamma_{\bs_2}\ldots\Gamma_{\bs_{k}})$, the same is true for $\boldBs$.

In order to turn $\objQ{k}$'s into $\objP{k}$'s, and thus, obtain the valid surrogate field, the sequences of arguments $\boldPrm = (\prm_1,\ldots,\prm_k)$ and $\boldBs=(\bs_1,\ldots,\bs_k)$ have to be forced to match up exactly. The first elements of the sequences match up by default because the classical trajectories corresponding to each Schr\"{o}dinger chain, end in the same point. If the beginning points $x_0$ and $y_0$ would be the same as well, then the classical trajectories would overlap and, as a result, the sequences would overlap too. The initial positions of each trajectory are determined by the initial state $\hat\rho_E$. Therefore, when the least action approximation applies, the environment facilitates the objective surrogate field representation when its initial state satisfies
\begin{align}
\langle x_0|\hat\rho_E|y_0\rangle = \delta(x_0-y_0)\rho_E(x_0).
\end{align}
Physically, this means that the environment should not be initialized in the Schr\"{o}dinger's cat type of state.

\section{Discussion}\label{sec:discussion}

\subsection{Impostor field representations}\label{sec:impostor}

Suppose that the system--environment arrangement is such that the interaction picture of the system density matrix is of the form analogous to Eq.~\eqref{eq:rhoS_stochastic}
\begin{align}
\label{eq:impostor_candidate}
\hat\rho_S^{I}(t)&= \sum_{k=0}^\infty (-i)^k\int_0^t\!\!dt_1\ldots\int_0^{t_{k-1}}\!\!\!dt_k\,
    F^{(k)}(\mathbf{t})
	\left(\prod_{l=1}^k\mathscr{V}_S(t_l)\right)\hat\rho_S.
\end{align}
If the family of multivariate functions $\{ F^{(k)} \}_{k=1}^\infty$ could be identified with {\it moments} of certain stochastic process, i.e., $F^{(k)}(\mathbf{t}) = \overline{\phi(t_1)\ldots\phi(t_k)}$, then it would follow that the dynamics of the system, at least in the case of this specific $SE$ arrangement, are simulable with a stochastic model $\hat H_\phi(t) = \hat H_S + \phi(t)\hat V_S$~\cite{Gu_JChemPhys2019}. However, it should be noted that, in general, the form~\eqref{eq:impostor_candidate} of the system state alone does not guarantee the existence of process $\phi(t)$ with moments fitting the corresponding functions $F^{(k)}$. Moreover, even if such process does exist, there is no general purpose systematic method for constructing the process given the family $\{F^{(k)}\}_{k=1}^\infty$; essentially, in order to identify $\phi(t)$, one has be able to recognize in $F^{(k)}$'s moments of known stochastic process. The one important exception is when functions $F^{(k)}$ follow the factorization pattern characteristic to Gaussian processes where the functions of order higher than $2$ factorize into specific combinations of $F^{(1)}$ and $F^{(2)}$, e.g., $F^{(4)}(\mathbf{t}) = F^{(2)}(t_1,t_2)F^{(2)}(t_3,t_4)+F^{(2)}(t_1,t_3)F^{(2)}(t_2,t_4) + F^{(2)}(t_1,t_4)F^{(2)}(t_2,t_3)$ (assuming that $F^{(1)}=0$). In such a case, there are only two functions to be fitted, and thus, it can be shown that it is always possible to find the matching Gaussian $\phi(t)$. It is vital to recognize, however, that such a factorization pattern is the unique property of Gaussian processes---there cannot exist a kind of ``super-Gaussian'' stochastic process, where all of its moments are expressed by a finite, but greater than two, number of independent autocorrelation functions~\cite{Marcinkiewicz_MZ39}.

Although, the ``fitted'' process described above and the objective surrogate field are both stochastic simulators, the ways they are established are very much different. Indeed, instead of looking for the best fit to the given (infinite) set of potential moments $F^{(k)}$, the surrogate field is constructed algorithmically from the ground up using $\hat H_E$, $\hat V_E$ and $\hat\rho_E$ that characterize the dynamical laws of the environment [see Eq.~\eqref{eq:P_SE}]. For the sake of clarity, we will label the fitted stochastic model according to the following 
\begin{definition}[Impostor field representation]
Any stochastic model $\hat H_\phi(t)$ that is {\it not} explicitly constructed by the means of the family of joint probability distributions $\{\objP{k}\}_{k=1}^\infty$ where $\objQ{k}(\boldPrm\boldBs\mathbf{t})\approx \delta_{\boldPrm,\boldBs}\objP{k}(\boldPrm\mathbf{t})$, but instead, is postulated or constructed in any other way under the constraint that its moments match certain form, will be referred to as an {\it impostor field} representation.
\end{definition}

The difference between surrogate and impostor representations extends beyond the way they are established. The defining feature of objective surrogate field representation is its inter-subjectivity in all contexts. On the other hand, the impostor representation is inherently subjective---i.e., it is context-dependent and is not necessarily valid in all contexts---because it is based on functions~$F^{(k)}$ that, in general, combine contributions from both sides of $SE$ arrangement. Hence, when the impostor is found, it can be used to simulate the decoherence caused by $E$ in the given specific context~\cite{Gu_JChemPhys2019}, but nothing beyond that purpose. If the impostor proves to be inter-subjective anyway, it can only be by accident, e.g., because it happens to be identical with the objective surrogate. We will illustrate these points with an example of {\it dephasing} qubit context defined by $\hat H_S=\omega(|{+}\rangle\langle{+}|-|{-}\rangle\langle{-}|)/2 = \omega\hat\sigma_z/2$ (with $\omega=0$ for simplicity) and $\hat V_S = \hat\sigma_z/2$. The special feature of this particular context is that the density matrix of the system always has the form~\eqref{eq:impostor_candidate} and functions $F^{(k)}$ can be expressed in terms of joint quasi-probabilities
\begin{align}
\nonumber
F_{\mathrm{qubit}}^{(k)}(\boldT) &= \sum_{\boldPrm,\boldBs\in\Omega_{\hat V}^{\times k}}\!\delta_{\prm_1,\bs_1}
	\left(\prod_{l=1}^k\frac{\prm_l+\bs_l}2\right)\objQ{k}(\boldPrm\boldBs\boldT)\\
&= \sum_{\bm{\phi}\in\Omega_{\{V\}}^{\times k}}\left(\prod_{l=1}^k\phi_l\right)
	\underbrace{
		\Bigg(\prod_{l=1}^k\,\sum_{\substack{\prm_l,\bs_l:\,2\phi_l = \prm_l+\bs_l}}\Bigg)\delta_{\prm_1,\bs_1}\objQ{k}(\boldPrm\boldBs\boldT)
	}_{\equiv f^{(k)}(\bm{\phi}\boldT)},
\end{align}
where $\Omega_{\{V\}}$ contains all unique values $\phi = (\prm + \bs)/2$ and $\prm,\bs\in\Omega_{\hat V}$. Formally, the members of $\{f^{(k)}\}_{k=1}^\infty$ are sub-ensembles of joint quasi-probabilities, and thus, they inherit from $\{ \objQ{k} \}_{k=1}^\infty$ the compliance with the Chapman-Kolmogorov consistency criterion. Therefore, $f^{(k)}(\bm{\phi}\boldT)$'s also count as a joint quasi-probability distributions, but of quantum process $\phi(t)$, instead of $(\prm(t),\bs(t))$. For functions $F^{(k)}_\mathrm{qubit}$ to be identified with moments of stochastic process, however, $f^{(k)}$'s have to be downgraded to proper joint probability distributions, which means that they have to be non-negative
\begin{align}
\label{eq:QIDAS_impostor_condition}
 f^{(k)}(\bm{\phi}\boldT)\geqslant 0 \ \Leftrightarrow\ 
 	 \Bigg(\prod_{l=1}^k\,\sum_{\prm_l,\bs_l :\, 2\phi_l=\prm_l+\bs_l}\Bigg)
 	 \delta_{\prm_1,\bs_1}\Delta\objQ{k}(\boldPrm\boldBs\boldT) \geqslant 0.
\end{align}
Of course, when the environment satisfies criterion~\ref{crt:obj} (so that $\Delta\objQ{k} \approx 0$), the above condition is met, and stochastic process $\phi(t)$ defined by $\{f^{(k)}\}_{k=1}^\infty$ is the same as the objective surrogate field. However, the form of condition~\eqref{eq:QIDAS_impostor_condition} allows for another possibility: $f^{(k)}$'s can be non-negative even when $\Delta\objQ{k}$'s are non-negligible, e.g., because of constructive interference between the constituting propagator chains. When this is the case, the impostor representation $\phi(t)$ exists while the surrogate representation is invalid. In practical terms, this means that, even though, the dynamics of the dephasing qubit can be described with model $\hat H_\phi(t) = \phi(t)\hat\sigma_z/2$, the stochastic simulation would break down when the qubit is swapped for a different system. An example of such a scenario was observed in Refs.~\cite{Paz-Silva_PRA2017,Kwiatkowski_PRB2020} where it was demonstrated that the Gaussian stochastic model fitted to $F^{(k)}$ obtained for dephasing qubit with $\hat V_S = \hat\sigma_z/2$ is no longer valid when the coupling is swapped with $\hat V_S = (\hat{\mathds{1}} + \hat\sigma_z)/2 = |{+}\rangle\langle{+}|$. In fact, unless the environment facilitates its surrogate field, this seemingly insignificant change of context renders impostor representations impossible because the new coupling operator causes the system state to deviate from the form~\eqref{eq:impostor_candidate} by introducing the imaginary part to the second moment~\cite{Gu_JChemPhys2019}.

\subsection{What is classical about surrogate field?}\label{sec:loc_real}

In classical theory, a particle is considered an {\it element of objective reality}---it is assumed that it unconditionally exists in some definite state at all times. In the formalism of the theory, the state of the particle is equated to continuous single-valued trajectory $\mathbf{r}(t)$ representing the position of its center of mass as a function of time. If the system is composed of multiple particles labeled with index $i$, the description is extended by simply including a trajectory $\mathbf{r}_i(t)$ for each constituent so that each one of them is an element of objective reality.

Note that the unconditional existence assumption implies that the state of classical particle is inter-subjective. Indeed, since the position and the momentum are definite at all times, then all observers will report the same result when they measure them at the given moment in time. This points to the first analogy between classical theory and the surrogate field representation. When we know that any system coupled to the environment that facilitates its objective surrogate will experience the same field, and that the experience of such systems is the only possible record about the surrogate, then it makes no practical difference if we choose to presume that the surrogate exists even if no one is ``looking''. Therefore, we can say that the objective surrogate field can be considered an element of objective reality.

Although the very fact of the classical particle's existence---formally represented by uninterrupted generation of its trajectory---does not rely on any other agent, these ``other agents'' can intervene and cause the particle's trajectory to be modified. In the formalism of the theory, the modifications due to particles' interactions are governed by an appropriate set of coupled equations of motion for all trajectories. However, it is impossible to store an unambiguous record about the form of equations of motion in any of those modified trajectories. Or in other words, the same set of trajectories could result from whole plethora of different sets of equations. In particular, it is always possible to replace equations that couple many trajectories through interaction potentials with a set of decoupled equations where each particle experiences an {\it external force field}. Equivalently, one can describe the dynamics of these particles in terms of constrained motion---the method that allows to ``conceal'' most of (or even all) such force fields by switching to properly chosen set of generalized coordinates. Hence, one can always describe a multi-particle system in terms of independent particles, each riding on an elaborately constructed track that leads it over trajectory that is identical to one generated in the presence of interactions. The model of epicycles in Ptolemaic system of astronomy is an example of such an approach. 

The concept of external force fields and the method of constrained motion, naturally supported by classical theory, are, in general, not compatible with the formalism of quantum mechanics. However, the cases when the surrogate field, or even the impostor fields, are valid, represent exceptions when a multi-party quantum system allows this kind of semi-classical description. It is the second reason why surrogate field representation can be considered classical.

\subsection{Surrogate field and back-action}\label{sec:back-action}

A commonly entertained hypothesis (e.g., see Refs.~\cite{DasSarma_PRB2014,Ma_PRB2015,Zhao_PRL2011,Bethke_arXive2019}) proposes that for the stochastic modeling of system--environment interaction to work, the coupling between $S$ and $E$ has to cause no {\it back-action}. The absence of back-action is understood here as the asymmetry between the system and the environment where $E$ influences $S$ but $S$ does not influence $E$. 

This hypothesis can be motivated by the following intuitive reasoning. When there is no back-action, it stands to reason that $E$ evolves as if $S$ did not exist, and hence, the environment can always be assigned with a definite state $\hat\rho_E(t)$ as the dynamical equation of its motion is decoupled from the system. Moreover, if the state of one of the parties is definite at all times, then the state of the total system can only be separable
\begin{align}
\hat\rho_{SE}(t) &\sim \hat\rho_S\left (t| \{\hat\rho_E(\tau): 0<\tau<t\}\right)\otimes\hat\rho_E(t),
\end{align}
where the evolution of the system state is, in general, dependent on the history of the environment [compare with Born approximation~\eqref{eq:Born} of Sec.~\ref{sec:open}.] When this is the case, it seems reasonable to anticipate that, from the point of view of the system, $E$ would act as a source of external (i.e., independent of $S$) field that drives its evolution. On the other hand, if the system evolves as if driven by an external field, it seems self evident that it would be a contradiction if $S$ was able to influence the field's source. In what follows, we will investigate if this line of argument holds up.

The back-action will be considered absent (or, at least, negligible) when the expectation value of any $E$-only observable is unchanged in comparison to the value obtained in the case when there is no system--environment coupling. Formally, this criterion is expressed as
\begin{align}
\nonumber
&\mathrm{tr}_S\left(
	e^{-i t(\hat H_S+\hat H_E+\hat V_S\otimes\hat V_E)}
	\hat\rho_S\otimes\hat\rho_E
	\,e^{i t(\hat H_S+\hat H_E+\hat V_S\otimes\hat V_E)}\right)\\
\label{eq:no-back-action}
&\phantom{=}\approx
	\mathrm{tr}_S\left(e^{-i t(\hat H_S+\hat H_E)}
	\hat\rho_S\otimes\hat\rho_E
	\,e^{i t(\hat H_S+\hat H_E)}\right),
\end{align}
where $\mathrm{tr}_S$ indicates the partial trace over system degrees of freedom. With the use of this criterion, and the following counter examples, we will now show that the argument presented above is faulty and that there is no {\it causal} link between the lack of back-action and the validity of surrogate field representation.

First, we choose $S$ to be a dephasing qubit system with $\hat H_S = 0$ and $\hat V_S = (|{+}\rangle\langle{+}|-|{-}\rangle\langle{-}|)/2 = \hat\sigma_z/2$ and an arbitrary $\hat H_E$, $\hat V_E$. Then, when the interaction is present, with some algebra, we can express the reduced density matrix of the environment in the terms of propagator chains
\begin{align}
\nonumber
&\mathrm{tr}_S\left(
	e^{-it(\hat H_E+\frac{1}{2}\hat\sigma_z\otimes\hat V_E)}
	\hat\rho_S\otimes\hat\rho_E
	e^{it(\hat H_E+\frac{1}{2}\hat\sigma_z\otimes\hat V_E)}
	\right)
	=\sum_{s=\pm}\langle s|\hat\rho_S|s\rangle 
	e^{-i t(\hat H_E + \frac{1}{2}s\hat V_E)}
	\hat\rho_E
	e^{i t(\hat H_E + \frac{1}{2}s\hat V_E)}\\
\nonumber
&\phantom{=}
	=\sum_{s=\pm}\langle s|\hat\rho_S|s\rangle\sum_{k=0}^\infty \left(-i\frac{s}{2}\right)^k
	\int_0^t \!dt_1\ldots\int_0^{t_{k-1}}\!\!\!dt_k\,
	\Bigg(\prod_{l=1}^k\sum_{\substack{n_l\neq m_l}}\Bigg)\Bigg[\prod_{l=1}^k (v_{n_l}-v_{m_l})\Bigg]\\
\label{eq:rhoE_with_int}
&\phantom{=\times}\times
	 e^{-i (t - t_1)\hat H_E}|n_1\rangle\langle m_1|e^{i (t - t_1)\hat H_E} 
	\left(\prod_{l=1}^{k-1} T_{t_l-t_{l+1}}(n_lm_l|n_{l+1}m_{l+1})\right)
	\langle n_k|\hat\rho_E(t_k)|m_k\rangle.
\end{align}
Note that the links in the chains are connected only through coherences (i.e., the index pairs in each sum cannot match up). With this in mind, we take the initial state $\hat\varrho_E \propto \hat{\mathds{1}}$. Then, in the above expression, only $k=0$ term survives because the initial link of each chain vanishes as $\langle{n_k}|e^{-i t_k \hat H_E}\hat{\mathds{1}}e^{i t_k\hat H_E}|m_k\rangle = 0$, which leaves us with
\begin{align}
\nonumber
&\mathrm{tr}_S\Big(
	e^{-i t (\hat H_E+\frac{1}{2}\hat\sigma_z\otimes\hat V_E)}
	\overbrace{\hat\rho_S\otimes\hat\rho_E}^{\propto\hat\rho_S\otimes\hat{\mathds{1}}}
	e^{i t (\hat H_E+\frac{1}{2}\hat\sigma_z\otimes\hat V_E)}
	\Big)
	=\sum_{s=\pm}\langle s|\hat\rho_S|s\rangle \, e^{-it\hat H_E}\hat\rho_E\,e^{it\hat H_E}\\
&\phantom{=}
 	=\mathrm{tr}_S\Big(e^{-it\hat H_E}\hat\rho_S\otimes\hat\rho_E\, e^{it\hat H_E}\Big)
	=\mathrm{tr}_S\Big(
 		e^{-it(\hat H_S+\hat H_E)}
 		\hat\rho_S\otimes\hat\rho_E
 		\,e^{-it(\hat H_S+\hat H_E)}
 		\Big),
\end{align}
i.e., according to criterion \eqref{eq:no-back-action}, the back-action disappears. On the other hand, restricting the form of the initial state is not sufficient to ensure that each $\Delta\objQ{k}\approx 0$. Therefore, even though there is no back-action, the surrogate field representation is not guaranteed to be valid. 

With another counter example, we can also disprove the reciprocal assertion that a valid surrogate field implies the lack of back-action. In this case, we keep the same choice for $S$ as before, but we specify $E$ to be of the quasi-static coupling type discussed in Sec.~\ref{sec:quasi-static}. When $[\hat H_E,\hat V_E]=0$, the expressions for the reduced density matrix with, and without, system--environment coupling simplify as follows
\begin{align}
\nonumber
&\mathrm{tr}_S\Big(
	e^{-i t (\hat H_E+\frac{1}{2}\hat\sigma_z\otimes\hat V_E)}
	\hat\rho_S\otimes\hat\rho_E
	\, e^{i t (\hat H_E+\frac{1}{2}\hat\sigma_z\otimes\hat V_E)}
	\Big)\\
\label{eq:back-action_with_int}
&\phantom{=}
	=\sum_{n,m}|n\rangle\langle m|e^{i t(\epsilon_m - \epsilon_n)}\langle m|\hat\rho_E|n\rangle
	\sum_{s=\pm}\langle s|\hat\rho_S|s\rangle
	e^{\frac{i}{2} s(v_m - v_n)},\\
\label{eq:back-action_no_int}
&\mathrm{tr}_S\Big(e^{-i t\hat H_E}\hat\rho_S\otimes\hat\rho_E\, e^{it\hat H_E}\Big)
	= \sum_{n,m}|n\rangle\langle m|e^{i t(\epsilon_m-\epsilon_n)}\langle m|\hat\rho_E|n\rangle .
\end{align}
where $\hat H_E|n\rangle = \epsilon_n|n\rangle$. On the one hand, we have demonstrated in Sec.~\ref{sec:quasi-static} that quasi-static coupling facilitates valid surrogate field representation. On the other hand, by comparing Eqs.~\eqref{eq:back-action_with_int} and \eqref{eq:back-action_no_int} we can see that, in those same circumstances, the qubit can still influence the environment. Hence, it is possible that a valid surrogate field representation exists, while $S$ exerts the back-action onto~$E$.

The above examples demonstrate that, contrary to the ``common sense'' intuition, there is no causal link between the lack of back-action and surrogate field representation. We believe the reason for this counter-intuitive disconnect can be explained with another intuitive picture. As we argued previously, no back-action means that the state of $E$ remains definite and is independent of $S$, hence, it is a statement about the environment as a whole. On the other hand, the surrogate field representation de-emphasizes the role of the state of the environment $\hat\rho_E(t)$, and instead, places the focus on the coupling $\hat V_E(t)$ and its dynamics: when the valid surrogate exists, one could say that it is the ``state'' of the coupling operator which remains definite (or that it can be assigned with a definite ``value''), and so it can be as well superseded with an external field. As it turns out, the way the state of the environment evolves is not necessarily the decisive factor in determining the ``state'' of the coupling.

\subsection{Surrogate field and system--environment entanglement}\label{sec:entanglement}

When the initial states $\hat\rho_S$ and $\hat\rho_E$ are pure, it is known~\cite{Schlosshauer_book}, that any subsequent loss of purity in $S$ is caused by the formation of entanglement between the system and the environment. When this is the case, it is said that the system undergoes the process of {\it quantum} decoherence. When the causes for the loss of purity are of secondary importance, or they cannot be unambiguously identified with the entanglement formation, it is said that $S$ undergoes  {\it wide-sense} decoherence, or simply decoherence without adjectives (as we are using it throughout the paper). It has been demonstrated~\cite{Eisert_PRL02,Hilt_PRA09,Maziero_PRA10,Pernice_PRA11,Roszak_PRA2015,Roszak_PRA18}, that in a more realistic case of mixed $\hat\rho_E$, the correlation between the presence (or amount) of $SE$ entanglement, and the severity of the purity decay in $S$ is rather weak or even nonexistent, e.g., in the process of pure dephasing of a qubit coupled with $E$ initialized in maximally mixed state,  $\hat\rho_E\propto\hat{\mathds{1}}$, the $SE$ entanglement can never form~\cite{Roszak_PRA2015}. Therefore, in general, the question of the relationship between decoherence and $SE$ entanglement is rather uninteresting. However, given that the objective surrogate field can be considered classical (see Sec.~\ref{sec:loc_real}), one might be tempted to surmise that its source, the environment itself, is also effectively a classical system that, by its very nature, is unable to participate in quantum correlation such as entanglement. In other words, one might presume that the formation of $SE$ entanglement is incompatible with surrogate field. This line of reasoning leads to a strong suggestion that, in this specific case, there is a link between entanglement (or rather, its absence) and the surrogate field-induced decoherence.

We will show now that such a conclusion is incorrect, and there is no causal link between the lack of entanglement and the validity of surrogate representation. For this purpose, let us utilize the recently discovered criterion for the absence of entanglement between dephasing qubit (i.e., $\hat H_S=0$ and $\hat V_S = \hat\sigma_z/2$) and its environment \cite{Roszak_PRA2015} which reads: the dephasing qubit is {\it not} entangled with its environment at time $t$ if and only if
\begin{align}
\label{eq:QEE}
&e^{-it(\hat H_E + \frac{1}{2}\hat V_E)}\hat\rho_E\, e^{it(\hat H_E + \frac{1}{2}\hat V_E)}
    =e^{-it(\hat H_E - \frac{1}{2}\hat V_E)}\hat\rho_E\,e^{it(\hat H_E - \frac{1}{2}\hat V_E)}.
\end{align}

First, set the initial state of $E$ to $\hat\rho_E \propto \hat{\mathds{1}}$, then Eq.~\eqref{eq:QEE} is trivially satisfied. On the other hand, specifying the initial state of $E$ is not sufficient for ensuring the validity of the surrogate representation. Therefore, even though there is no entanglement with the environment, the surrogate field might not exist.

Second, choose $E$ to be of the quasi-static coupling type (i.e., $[\hat H_E,\hat V_E]=0$), then the surrogate field representation is guaranteed to be valid, but the criterion \eqref{eq:QEE} is not necessarily satisfied because
\begin{align}
&e^{-it(\hat H_E \pm \frac{1}{2}\hat V_E)}\hat\rho_E e^{it(\hat H_E \pm \frac{1}{2}\hat V_E)}
	= \sum_{n,m}|n\rangle\langle m| e^{it[(\epsilon_m-\epsilon_n) \pm \frac{1}{2}(v_m - v_n)]}\langle n|\hat\rho_E|m\rangle,
\end{align}
which shows that the l.h.s of Eq.~\eqref{eq:QEE} differs from the r.h.s unless the initial state is diagonal in $\{|n\rangle\}_n$ basis. Hence, even when the surrogate field representation is valid, the system can still become entangled with its environment.

As it was the case in the previous section, also here, the causal link has to be dismissed. The reasons for the disconnect are essentially the same as before: the entanglement is a statement about the state of system--environment complex, while the surrogate field representation is concerned only with substituting for the coupling.

\subsection{Multi-component surrogate field}\label{sec:mulit-component}

In general, the system--environment coupling has a form of a compound operator
\begin{align}
\hat H_{SE} = \hat H_S\otimes\hat{\mathds{1}}+\hat{\mathds{1}}\otimes\hat H_E + \sum_{\lambda = 1}^\Lambda \hat V_S^\lambda\otimes\hat V_E^\lambda,
\end{align}
with Hermitian constituents $\hat V_S^\lambda$ and $\hat V_E^\lambda$. This also includes the ``non-Hermitian'' couplings $\hat a_S\otimes \hat a_E + \hat a_S^\dagger\otimes \hat a_E^\dagger$, because they can always be written as a compound operator with $\hat V_{S/E}^{1} = (\hat a_{S/E}+\hat a_{S/E}^\dagger)/\sqrt 2$ and $\hat V_{S/E}^2 = \pm i(\hat a_{S/E}-\hat a_{S/E}^\dagger)/\sqrt 2$.

A valid surrogate field representation for such a coupling utilizes the model
\begin{align}
\hat H_{\Xi}(t) = \hat H_S + \sum_{\lambda=1}^\Lambda \Xi^\lambda(t)\hat V_S^\lambda,
\end{align}
where the surrogate is a multi-component stochastic process $\Xi(t) = (\Xi^1(t),\ldots,\Xi^\Lambda(t))$ governed by the family of joint probability distributions $P_\Xi^{(k)}(\bm{\Xi}\boldT)$. However, the evolution is not directly determined by these joint probabilities, as we have
\begin{align}
\nonumber
    \hat\rho^I_S(t) &= e^{it\hat H_S}\overline{\hat U(t|\Xi)\hat\rho_S\hat U^\dagger(t|\Xi)}e^{-it\hat H_S}\\
\label{eq:multi_surrogate}
    &=\sum_{k=0}^\infty(-i)^k\int_0^t\!\!dt_1\ldots\int_0^{t_{k-1}}\!\!dt_k
    \left(\prod_{l=1}^k\sum_{\lambda_l=1}^\Lambda\sum_{\prm_l\in\Omega_{\Xi^{\lambda_l}}}\right)
        P^{(k)}_{\lambda_1\ldots\lambda_k}(\boldPrm\boldT)
        \left(\prod_{l=1}^k\prm_l\mathscr{V}_S^{\lambda_l}(t_l)\right)\hat\rho_S,
\end{align}
where $\mathscr{V}_S^\lambda(t)\hat A = \hat V_S^\lambda(t)\hat A - \hat A\hat V_S^\lambda(t)$ and
\begin{align}
    P_{\lambda_1\ldots\lambda_k}^{(k)}(\boldPrm\boldT) =
        \sum_{\bm{\Xi}\in\Omega_\Xi^{\times k}}\left(\prod_{l=1}^k \delta_{\prm_l , \Xi^{\lambda_l}_l}\right)
        P^{(k)}_\Xi(\bm{\Xi}\boldT)
\end{align}
are the \emph{marginal} joint distributions. As sub-ensembles of $P_\Xi^{(k)}$'s, these distributions satisfy the non-negativity and consistency criteria
\begin{align}
&\sum_{\prm_l\in\Omega_{\Xi^{\lambda_l}}}P^{(k)}_{\lambda_1\ldots\lambda_k}(\boldPrm\boldT)
        =P^{(k-1)}_{\ldots\lambda_{l-1}\lambda_{l+1}\ldots}(\ldots;\prm_{l-1}t_{l-1};\prm_{l+1}t_{l+1};\ldots).
\end{align}

In the case of the exact Hamiltonian, given the spectral decomposition of operators in the compound coupling
\begin{align}
    \hat V_E^\lambda =\sum_n v_{n;\lambda}|n;\lambda\rangle\langle n;\lambda|
    = \sum_{\prm\in\Omega_{\hat V^\lambda}}\prm\sum_{n:\prm = v_{n;\lambda}}|n;\lambda\rangle\langle n;\lambda|,
\end{align}
the evolution of the reduced system state reads
\begin{align}
\nonumber
    \hat\rho^I_S(t) &= e^{it\hat H_S}\mathrm{tr}_E\left(e^{-it\hat H_{SE}}\hat\rho_S e^{it\hat H_{SE}}\right)e^{-it\hat H_S}\\
\nonumber
    &= \sum_{k=1}^\infty(-i)^k\int_0^t\!\!dt_1\ldots\int_0^{t_{k-1}}\!\!dt_k
        \left(\prod_{l=1}^k\sum_{\lambda_l=1}^\Lambda\sum_{\prm_l,\bs_l\in\Omega_{\hat V^{\lambda_l}}}\right)\\
&\phantom{=}\times
        \delta_{\prm_1,\bs_1}q^{(k)}_{\lambda_1\ldots\lambda_k}(\boldPrm\boldBs\boldT)
        \left(\prod_{l=1}^k\mathscr{W}_S^{\lambda_l}(\prm_l\bs_lt_l)\right)\hat\rho_S,
\end{align}
where $\mathscr{W}_S^\lambda$ are defined analogously to $\mathscr{W}_S$ [see Eq.~\eqref{eq:W_superop}] but with $\hat V_S^\lambda(t)$ replacing $\hat V_S(t)$. The quasi-probabilities corresponding to the marginal distributions in~\eqref{eq:multi_surrogate} are given by
\begin{align}
\nonumber
    q_{\lambda_1\ldots\lambda_k}^{(k)}(\boldPrm\boldBs\boldT) &=
    \Bigg(\prod_{l=1}^k\sum_{\substack{n_l:\\ \prm_l = v_{n_l;\lambda_l}}}\sum_{\substack{m_l:\\\bs_l = v_{m_l;\lambda_l}}}\Bigg)
    \delta_{n_1,m_1}\\
    &\phantom{=}\times
        \langle n_k;\lambda_k|\hat\rho_E(t_k)|m_k;\lambda_k\rangle\prod_{l=1}^k T_{t_l-t_{l+1}}^{\lambda_l\lambda_{l+1}}(n_lm_l|n_{l+1}m_{l+1}),
\end{align}
where the propagators have been modified according to
\begin{align}
    &T^{\lambda\lambda'}_t(nm|n'm')= \mathrm{tr}_E\left(|m;\lambda\rangle\langle n;\lambda|e^{-it\hat H_E}|n';\lambda'\rangle\langle m';\lambda'|e^{it\hat H_E}\right).
\end{align}
These marginal quasi-probabilities are consistent
\begin{align}
    &\sum_{\prm_l,\bs_l\in\Omega_{\hat V^{\lambda_l}}}q_{\lambda_1\ldots\lambda_k}^{(k)}(\boldPrm\boldBs\boldT)
    = q_{\ldots\lambda_{l-1}\lambda_{l+1}}^{(k-1)}(\ldots;\prm_{l-1}\bs_{l-1}t_{l-1};\prm_{l+1}\bs_{l+1}t_{l+1};\ldots),
\end{align}
and, due to the contribution from coherence-connected chains, they are not necessarily non-negative. Therefore, the validity criterion for multi-component surrogate field representation is virtually identical to criterion~\ref{crt:obj}: when the superposition of coherence-connected propagator chains in each quasi-probability is negligible, then the remaining non-negative projector-connected chains $p_{\lambda_1\ldots\lambda_k}^{(k)}(\boldPrm\boldT)$ can be treated as a proper marginal joint probability distributions. When this is the case, then the evolution of the reduced state of any $S$ coupled to $E$ through operator compounded from any combination of $\hat V_E^\lambda$'s is indistinguishable from the simulation with multi-component surrogate field.

However, the issue is that this criterion is only an existence theorem: when it is satisfied, we only know that the objective surrogate $\Xi(t)$ exists and that the stochastic simulation is valid, but we cannot access the multi-component trajectories of the surrogate to run this simulation with. Indeed, the projector-connected chains $p_{\lambda_1\ldots\lambda_k}^{(k)}$ are only marginal distributions, and hence, even when one calculates all of them, it is still not enough information to recover the distributions they are marginalizing---the family analogous to $\{P_\Xi^{(k)}\}_{k=1}^\infty$ that is needed to instantiate trajectories. The exception is when the components of the surrogate are mutually independent, which occurs when $\hat V_E^\lambda$'s couple to separate sub-environment, i.e., when $\hat\rho_E = \bigotimes_\lambda\hat\rho^\lambda$, $\hat H_E = \sum_\lambda \hat{\mathds{1}}^{\otimes\lambda-1}\otimes\hat H^\lambda\otimes\hat{\mathds{1}}^{\otimes\Lambda-\lambda+1}$ and $\hat V_E^\lambda = \hat{\mathds{1}}^{\otimes\lambda-1}\otimes\hat V^\lambda\otimes\hat{\mathds{1}}^{\otimes\Lambda-\lambda+1}$, and each of those sub-environments facilitates its own surrogate field. Therefore, only in this case, the multi-component surrogate representation is useful in practical terms.

\section{Conclusion}

We have formulated the sufficient criterion for the dynamics of any open quantum system coupled with a given environment to be simulated using the external field that is a surrogate for the environmental degrees of freedom---the surrogate field representation. To achieve this, we have developed the approach in which the influence of the environment is wholly described by the family of joint quasi-probabilities $\{\objQ{k}\}_{k=1}^\infty$, with each of its members constructed out of simple basic elements. This language has proven to be flexible enough to allow us not only to carry out a comprehensive analysis of microscopic origins of so-called classical noise approximations and random unitary dynamical maps, but also to explore some of the most interesting accompanying issues. Two important examples of such issues were the previously hypothesized incompatibility of surrogate representation with the formation of system--environment entanglement, and the causal relation between the absence of system's back-action and the existence of valid surrogate representation; we have disproved both propositions.

We have concluded that it is impossible to point to {\it one} reason for the validity of the surrogate field representation (like e.g., the absence of back-action). Instead, whether the simulation with surrogate field is valid is determined by the relationship between the dynamical laws governing the environment (the free Hamiltonian $\hat H_E$ and the initial state~$\hat\rho_E$) and operator $\hat V_E$ that couples it to the system. The examples of environment types that facilitate their surrogate fields presented here illustrate this point by showing a variety of ways to satisfy the validity criterion. 

We have addressed the issue of subjectivity and inter-subjectivity of the surrogate field representation. Even though the question of the objectivity of external field simulator is an important one---both from practical and purely theoretical point of view---previous studies on classical noise or random unitary maps were unable to engage with it in satisfactory capacity. We had taken this particular shortcoming into consideration, and we had set fixing this specific blind spot as one of the main design goals of our approach. The resultant quasi-probability formulation leads to the system state decomposition~\eqref{eq:rhoS_quantum} where the contributions from the system and the environment are clearly separated. This separation is crucial; it allows for the influence exerted by the environment to be considered independently of the influenced system (e.g., in order to determine whether this influence can be represented with the surrogate field). Thus, the quasi-probability formulation was an ideal tool for finding the answer to the question of surrogate's objectivity; one can hope that it will also open new avenues for the development of the quantum open systems theory.

\section*{Acknowledgments}
We would like to thank M.~Ku\'{s} and F.~Sakuldee for insightful and helpful discussions. This work is supported by funds of Polish National Science Center (NCN), grant no.~2015/\allowbreak 19/\allowbreak B/\allowbreak ST3/\allowbreak 03152.

\appendix

\section{Reduced system state}\label{sec:rhoS_quantum}
The interaction picture of the reduced state of the system is given by
\begin{align}
\hat\rho_S^I(t) &= e^{it\hat H_S}\mathrm{tr}_E\left(
	e^{-it\hat H_{SE}}\hat\rho_S\otimes\hat\rho_E\,e^{it\hat H_{SE}}
	\right)e^{-it\hat H_S}
=e^{it[\hat H_S,\bullet]}\mathrm{tr}_E(\bullet)e^{-it[\hat H_{SE},\bullet]}\hat\rho_S\otimes\hat\rho_E,
\end{align}
where we have switched to the super-operator representations,
\begin{subequations}
\label{eq:super-op_rep}
\begin{align}
&\mathrm{tr}_E(\bullet)\hat A\otimes\hat B = \mathrm{tr}(\hat B)\hat A,\\
&[\hat A,\bullet]\hat B = [\hat A,\hat B] = \hat A\hat B-\hat B\hat A,\\
&\left(e^{-it\hat A}\bullet e^{it\hat A}\right)\hat B = e^{-it \hat A}\hat B\,e^{it\hat A} = e^{-it[\hat A,\bullet]}\hat B.
\end{align}
\end{subequations}
Continuing,
\begin{align}
\nonumber
\hat\rho_S^I(t) &= e^{it[\hat H_S,\bullet]}\mathrm{tr}_E(\bullet)
	e^{-it[\hat H_S+\hat H_E+\hat V_S\otimes\hat V_E,\bullet]}
	\hat\rho_S\otimes\hat\rho_E\\
\nonumber
&=e^{it[\hat H_S,\bullet]}\mathrm{tr}_E(\bullet)
	e^{-it[\hat H_S,\bullet]}e^{-it[\hat H_E,\bullet]}\\
\nonumber
&\phantom{=}\times
	\mathcal{T}{\exp}\left(-i\int_0^t  e^{i\tau[\hat H_S+\hat H_E,\bullet]}
		[\hat V_S\otimes\hat V_E,\bullet]
		e^{-i\tau[\hat H_S+\hat H_E,\bullet]}d\tau\right)
		\hat\rho_S\otimes\hat\rho_E\\
\label{eq:full_rhoS}
&=\sum_{k=0}^\infty (-i)^k
	\int_0^t\!\!dt_1\ldots\int_0^{t_{k-1}}\!\!\!dt_k
	\,\mathrm{tr}_E(\bullet)
	\left(
		\prod_{l=1}^k e^{it_l[\hat H_E,\bullet]}[\hat V_S(t_l)\otimes\hat V_E,\bullet]e^{-it_l[\hat H_E,\bullet]}
	\right)\hat\rho_S\otimes\hat\rho_E
\end{align}
where symbol $\prod_{l=1}^{k} \mathscr{A}(l)$ applied to super-operators is understood as an ordered composition $\mathscr{A}(1)\mathscr{A}(2)\ldots \mathscr{A}(k)$, $\mathcal{T}$ indicates time-ordering operation, and $\hat V_S(t) = e^{it\hat H_S}\hat V_Se^{-it\hat H_S} = e^{it[\hat H_S,\bullet]}\hat V_S$. 

First, note the following identity for super-operator associated with commutator of a composite operator
\begin{align}
\label{eq:commutator_identity}
[\hat V_S(t)\otimes\hat V_E ,\bullet] = \frac{1}{2}[\hat V_S(t),\bullet]\otimes\{\hat V_E,\bullet\}
	+\frac{1}{2}\{\hat V_S(t),\bullet\}\otimes[\hat V_E,\bullet],
\end{align}
where $\{ \hat A,\hat B\}= \hat A\hat B+\hat B\hat A$ is the anti-commutator. Next, let $\{|n\rangle\}_n$ be the basis in $E$ composed of eigenstates of the environment-side coupling, $\hat V_E|n\rangle = v_n|n\rangle$. Then, the set $\{|n\rangle\langle m|\}_{n,m}$ composed of projectors $|n\rangle\langle n|$ and coherences $|n\rangle\langle m|$ ($n\neq m$) forms an orthonormal basis in the subspace of linear Hermitian operators acting in $E$. Moreover, the elements of this basis are also the right eigenoperators of super-operator associated with commutator and anti-commutator of $\hat V_E$,
\begin{align}
\frac{1}{2}(\hat V_E\bullet\pm\bullet\hat V_E)|n\rangle\langle m| = \frac{v_n \pm v_m}2 |n\rangle\langle m|.
\end{align}
Since these super-operators are Hermitian [with respect to the trace inner product $(\hat A|\hat B) = \mathrm{tr}(\hat A^\dagger \hat B)$], $|n\rangle\langle m|$ are also their left eigenoperators, and thus, the super-operators be subjected to the spectral decomposition
\begin{align}
\nonumber
\frac{1}{2}(\hat V_E\bullet \pm \bullet\hat V_E) &= \sum_{n,m}\frac{v_n\pm v_m}2 |n\rangle\langle m|\mathrm{tr}_E(|m\rangle\langle n|\bullet)\\
\label{eq:spectral_decomposition}
&=\sum_{\prm,\bs\in\Omega_{\hat V}}\frac{\prm\pm\bs}2\sum_{\substack{n:\\\prm = v_n}}\sum_{\substack{m:\\\bs=v_m}}
	|n\rangle\langle m|\mathrm{tr}_E(|m\rangle\langle n|\bullet).
\end{align}
Combining~\eqref{eq:commutator_identity} and \eqref{eq:spectral_decomposition} gives us
\begin{align}
\nonumber
[\hat V_S(t)\otimes\hat V_E,\bullet] &=\sum_{\prm,\bs\in\Omega_{\hat V}}
	\left(\frac{\prm+\bs}2[\hat V_S(t),\bullet]+\frac{\prm-\bs}2\{\hat V_S(t),\bullet\}\right)\\
\nonumber
&\phantom{=}
	\otimes \left(\sum_{\substack{n:\prm = v_n}}\sum_{\substack{m:\bs=v_m}}|n\rangle\langle m|\mathrm{tr}_E(|m\rangle\langle n|\bullet)\right)\\
&=\sum_{\prm,\bs\in\Omega_{\hat V}}\mathscr{W}_S(\prm\bs t)\otimes
	\left(\sum_{\substack{n:\prm = v_n}}\sum_{\substack{m:\bs=v_m}}|n\rangle\langle m|\mathrm{tr}_E(|m\rangle\langle n|\bullet)\right),
\end{align}
which we then substitute into Eq.~\eqref{eq:full_rhoS}
\begin{align}
\nonumber
\hat\rho_S^I(t) &=\sum_{k=0}^\infty (-i)^k\int_0^t\!\!dt_1\ldots\int_0^{t_{k-1}}\!\!\!dt_k\,\mathrm{tr}_E(\bullet)
		\sum_{\boldPrm,\boldBs\in\Omega_{\hat V}^{\times k}}
		\Bigg[\left(\prod_{l=1}^k \mathscr{W}_S(\prm_l\bs_lt_l)\right)\hat\rho_S\\
\nonumber
&\phantom{=}\otimes
	\Bigg(\prod_{l=1}^k \sum_{\substack{n_l:\\\prm_l=v_{n_l}}}\sum_{\substack{m_l:\\\bs_l = v_{m_l}}}
		e^{it_l\hat H_E}|n_l\rangle\langle m_l|e^{-it_l\hat H_E}
		\mathrm{tr}_E(|m_l\rangle\langle n_l|e^{-t_l\hat H_E}\bullet e^{it_l\hat H_E})\Bigg)\hat\rho_E\Bigg]\\
\nonumber
&=\sum_{k=0}^\infty (-i)^k\int_0^t\!\!dt_1\ldots\int_0^{t_{k-1}}\!\!\!dt_k
	\sum_{\boldPrm,\boldBs\in\Omega_{\hat V}^{\times k}}\Bigg[\left(\prod_{l=1}^k \mathscr{W}_S(\prm_l\bs_lt_l)\right)\hat\rho_S\\
\nonumber
&\phantom{=}\times
\Bigg(\prod_{l=1}^k \sum_{\substack{n_l:\\\prm_l=v_{n_l}}}\sum_{\substack{m_l:\\\bs_l = v_{m_l}}}\Bigg)
	\mathrm{tr}_E(|n_1\rangle\langle m_1|)\langle n_k|\hat\rho_E(t_k)|m_k\rangle
	\prod_{l=1}^{k-1}T_{t_l-t_{l+1}}(n_lm_l|n_{l+1}m_{l+1})\\
&=\sum_{k=0}^\infty (-i)^k\int_0^t\!\!dt_1\ldots\int_0^{t_{k-1}}\!\!\!dt_k
	\sum_{\boldPrm,\boldBs\in\Omega_{\hat V}^{\times k}}
	\delta_{\prm_1,\bs_1}\objQ{k}(\boldPrm\boldBs\boldT)\left(\prod_{l=1}^k \mathscr{W}_S(\prm_l\bs_lt_l)\right)\hat\rho_S.
\end{align}

\section{Consistency criterion for joint quasi-probabilities}\label{sec:quasi-CK_crit}

Using the definition of joint quasi-probability distribution~\eqref{eq:q_def}, we can write the left hand side of the consistency criterion~\eqref{eq:q_CK} as
\begin{align}
\nonumber
\sum_{\prm_l,\bs_l\in\Omega_{\hat V}}\objQ{k}(\boldPrm\boldBs\boldT) &=
 \sum_{\substack{n_1,m_1:\\\prm_1=v_{n_1}\\\bs_1=v_{m_1}}}\ldots
 \left(\sum_{\prm_l,\bs_l\in\Omega_{\hat V}}
 	\sum_{\substack{n_l:\\\prm_l = v_{n_l}}}\sum_{\substack{m_l:\\\bs_l=v_{m_l}}}
 	\right)
 \ldots\sum_{\substack{n_k,m_k:\\\prm_k = v_{n_k}\\\bs_k = v_{m_k}}}\\
 \label{eq:CK_lhs}
&\phantom{=}\times
	\delta_{n_1,m_1}\prod_{b=1}^{k-1}T_{t_b-t_{b+1}}(n_bn_b|n_{b+1}m_{b+1}) \langle n_k|\hat\rho_E(t_k)|m_k\rangle.
\end{align}
The sum over $\prm_l$ and $\bs_l$ effectively lifts all constraints from the sums over indices $n_l$ and $m_l$
\begin{align}
\sum_{\prm_l,\bs_l\in\Omega_{\hat V}}\sum_{\substack{n_l:\\\prm_l = v_{n_l}}}\sum_{\substack{m_l:\\\bs_l=v_{m_l}}} = \sum_{n_l,m_l}.
\end{align}
These indices form a connection between a two-link segment of the propagator chain; using the explicit definition of propagator~\eqref{eq:propagator_def} and the super-operator representation~\eqref{eq:super-op_rep} introduced in appendix~\ref{sec:rhoS_quantum}, we can preform the summation across this segment
\begin{align}
\nonumber
&\sum_{n_l,m_l}T_{t_{l-1}-t_l}(n_{l-1}m_{l-1}|n_lm_l)T_{t_l-t_{l+1}}(n_lm_l|n_{l+1}m_{l+1})\\
\nonumber
&\phantom{=}
	=\mathrm{tr}_E(|m_{l-1}\rangle\langle n_{l-1}|\bullet)\\
\nonumber
&\phantom{==}\times
		e^{-i(t_{l-1}-t_{l})[\hat H_E,\bullet]}
		\left(\sum_{n_l,m_l}|n_l\rangle\langle m_l|\mathrm{tr}_E(|m_l\rangle\langle n_l|\bullet)\right)
		e^{-i(t_l-t_{l+1})[\hat H_E,\bullet]}|n_{l+1}\rangle\langle m_{l+1}|\\
\nonumber
&\phantom{=}=
	\mathrm{tr}_E(|m_{l-1}\rangle\langle n_{l-1}|\bullet)e^{-i(t_{l-1}-t_l)[\hat H_E,\bullet]}e^{-i(t_l-t_{l+1})[\hat H_E,\bullet]}|n_{l+1}\rangle\langle m_{l+1}|\\
\label{eq:stich}
&\phantom{=}=
	\mathrm{tr}_E(|m_{l-1}\rangle\langle n_{l-1}|e^{-i(t_{l-1}-t_{l+1})[\hat H_E,\bullet]}|n_{l+1}\rangle\langle m_{l+1}|)= T_{t_{l-1}-t_{l+1}}(n_{l-1}m_{l-1}|n_{l+1}m_{l+1}),
\end{align}
where we have utilized the super-operator variant of the decomposition of identity (recall from appendix~\ref{sec:rhoS_quantum} that $\{|n\rangle\langle m|\}_{n,m}$ is an orthonormal basis)
\begin{align}
\sum_{n,m}|n\rangle\langle m|\mathrm{tr}(|m\rangle\langle n|\bullet) = \bullet.
\end{align}
The two-link segment is merged into one propagator that links $|n_{l+1}\rangle\langle m_{l+1}|$ and $|n_{l-1}\rangle\langle m_{l-1}|$ directly, and thus, $\objQ{k}(\boldPrm\boldBs\boldT)$ is reduced to $\objQ{k-1}(\ldots;\prm_{l-1}\bs_{l-1}t_{l-1};\prm_{l+1}\bs_{l+1}t_{l+1};\ldots)$.

\section{Joint quasiprobability distributions for open environment}\label{sec:Q_4_driven}

Consider an orthonormal basis in $ED$ subspace $\{ |n;i\rangle \}_{n,i}$, where $|n;i\rangle = |n\rangle\otimes|i\rangle$, $\{|i\rangle\}_i$ is an arbitrary basis in subspace $D$, and $|n\rangle$ are the eigenstates of $\hat V_E$. Using this basis and the Schr\"{o}dinger representation of propagators [see Eqs.~\eqref{eq:Schroedinger_propagators} and \eqref{eq:ED_unitary}]
\begin{align}
\nonumber
T_t(ni,mj|n'i',m'j') 
    &=\langle n;i|\hat U_{ED}(t)|n';i'\rangle \langle m;j|\hat U_{ED}(t)|m';j'\rangle^*\\
&  =\langle n;i|\hat U_{ED}(t)|n';i'\rangle \langle m';j'|\hat U_{ED}^\dagger(t)|m;j\rangle,
\end{align}
we will now rewrite the general definition \eqref{eq:q_def} of $\objQ{k}$
\begin{align}
\nonumber
&\objQ{k}(\boldPrm\boldBs\mathbf{t})=
	\sum_{\substack{n_1:\\\prm_1 = v_{n_1}}}\Bigg(\prod_{l=2}^k \sum_{\substack{n_l:\\\prm_l=v_{n_l}}}\sum_{\substack{m_l:\\\bs_l=v_{m_l}}}\Bigg)
	\sum_{i_1}\sum_{\substack{i_2\ldots i_k\\j_2\ldots j_k}}
	\left(\prod_{l=1}^{k-1}\langle n_l; i_l| \hat U_{ED}(t_l-t_{l+1})|n_{l+1};i_{l+1}\rangle\right)\\
\nonumber
&\phantom{=}
	\times\langle n_k; i_k|\hat U_{ED}(t_k)\hat\rho_E\otimes\hat\rho_D\hat U_{ED}(t_k)^\dagger|m_k;j_k\rangle
	\left(\prod_{l=1}^{k-1}\langle m_l; j_l| \hat U_{ED}(t_l-t_{l+1})|m_{l+1};j_{l+1}\rangle\right)^*\\
\nonumber
&=\sum_{\substack{n_1:\\\prm_1 = v_{n_1}}}\Bigg(\prod_{l=2}^k\sum_{\substack{n_l:\\\prm_l=v_{n_l}}}\sum_{\substack{m_l:\\\bs_l=v_{m_l}}}\Bigg)
		\Bigg\{\sum_{i_1}\langle n_1;i_1|
			\left[\prod_{l=2}^k \hat U_{ED}(t_{l-1}-t_{l})|n_l\rangle\langle n_l|\left(\sum_{i_l}|i_l\rangle\langle i_l|\right)\right]\\
&\phantom{=}
	\times\hat U_{ED}(t_k)\hat\rho_E\otimes\hat\rho_D\hat U_{ED}^\dagger(t_k)
	\left[\prod_{l=k}^2 \left(\sum_{j_l}|j_l\rangle\langle j_l|\right)|m_l\rangle\langle m_l|\hat U_{ED}^\dagger(t_{l-1}-t_{l})\right]|n_1;i_1\rangle\Bigg\},
\end{align}
where the symbol $\prod_{l=l_b}^{l_e}\hat A(l)$ is to be understood as an ordered composition: $\hat A(l_b)\hat A(l_b+1)\ldots\hat A(l_e)$ for $l_b < l_e$, or $\hat A(l_b)\hat A(l_b-1)\ldots \hat A(l_e)$ for $l_b > l_e$. Since the sums over $i_l$ and $j_l$ are not constraint in any way, we get $\sum_{i_1}\langle n_1;i_1|\bullet|n_1;i_1\rangle = \langle n_1|\mathrm{tr}_D(\bullet)|n_1\rangle$ and $\sum_{i_l}|i_l\rangle\langle i_l| = \hat{\mathds{1}}$, which leads to Eq.~\eqref{eq:driven_Q}.

\section{How to solve the system dynamics in surrogate field representation}\label{sec:tutorial}
First step is to choose a method for solving the dynamics of system $S$ for a given time-dependent external field, e.g., Euler's method for integrating von Neumann equation,
\begin{align}
\nonumber
&\frac{d}{dt}\hat\rho_S^{(j)}(t) = -i[\hat H_S+\prm^{(j)}(t)\hat V_S,\hat\rho^{(j)}_S(t)]\\
&\quad\rightarrow\quad \hat\rho^{(j)}_S(t+h) \approx \hat\rho^{(j)}_S(t)-i h [\hat H_S+\prm^{(j)}(t)\hat V_S,\hat\rho^{(j)}_S(t)]
\ \mathrm{and}\ \hat\rho_S^{(j)}(0) = \hat\rho_S,
\end{align}
for small time step $h$. Index $j$ indicates the solution obtained for $j$th trajectory $\prm^{(j)}(t)$ (real-valued function of time) of the surrogate field $\prm$ (stochastic process defined by the family $\{\objP{k}\}_{k=1}^\infty$). The chosen method defines the time grid for sampling the field's trajectories (see Sec.~\ref{sec:crit}); in this case, it is a basic uniform grid $\boldT_\mathrm{grd} = \{ M h = t, \ldots, 2h, h , 0 \}$ with sample trajectories
\begin{align}
\boldPrm^{(j)}_\mathrm{smp} = \{ \prm_1^{(j)},\ldots,\prm^{(j)}_{M},\prm_{M+1}^{(j)} \} = \{ \prm^{(j)}(Mh),\ldots,\prm^{(j)}(h),\prm^{(j)}(0) \}.
\end{align}
Each trajectory is instantiated by drawing it at random from the joint probability distribution
\begin{align}
\objP{M+1}(\boldPrm\boldT_\mathrm{grd}) = \left(\prod_{l=1}^{M+1}\sum_{\substack{n_l:\prm_l=v_{n_l}}}\right)
	\langle n_{M+1}|e^{-ih\hat H_E}\hat\rho_E e^{ih\hat H_E}|n_{M+1}\rangle
	\prod_{l=1}^{M}|\langle n_l|e^{-ih\hat H_E}|n_{l+1}\rangle|^2.
\end{align}
In this way one populates an ensemble of trajectories $\{ \boldPrm^{(j)}_\mathrm{smp} \}_{j=1}^{N_s}$ and calculates the corresponding ensemble of density matrices $\{ \hat \rho_S^{(j)}(Mh) \}_{j=1}^{N_s}$. The individual members of this ensemble are meaningless and their only purpose is to calculate the physically meaningful average,
\begin{align}
\nonumber
\hat\rho_S(t) &= \overline{\hat U(t|\prm)\hat\rho_S\hat U^\dagger(t|\prm)}
    =\lim_{N\to\infty}\frac{1}{N}\sum_{j=1}^N \hat U(t|\prm^{(j)})\hat\rho_S\hat U^\dagger(t|\prm^{(j)})\\
&\approx \frac{1}{N_s}\sum_{j=1}^{N_s} \hat U(t|\prm^{(j)})\hat\rho_S\hat U^\dagger(t|\prm^{(j)})
	\approx \frac{1}{N_s}\sum_{j=1}^{N_s} \hat \rho_S^{(j)}(Mh).
\end{align}
The larger the ensemble, the more accurate the approximation; typically, the sufficient ensemble size is $N_s\sim 10^3-10^4$.

\bibliography{surrogate}

\end{document}